# Model comparison from LIGO–Virgo data on GW170817's binary components and consequences for the merger remnant

# Model comparison from LIGO–Virgo data on GW170817's binary components and consequences for the merger remnant

B P Abbott[1], R Abbott[1], T D Abbott[2], S Abraham[3], F Acernese[4,5], K Ackley[6], C Adams[7], V B Adya[8], C Affeldt[9,10], M Agathos[11,12], K Agatsuma[13], N Aggarwal[14], O D Aguiar[15], L Aiello[16,17], A Ain[3], P Ajith[18], G Allen[19], A Allocca[20,21], M A Aloy[22], P A Altin[8], A Amato[23], S Anand[1], A Ananyeva[1], S B Anderson[1], W G Anderson[24], S V Angelova[25], S Antier[26], S Appert[1], K Arai[1], M C Araya[1], J S Areeda[27], M Arène[26], N Arnaud[28,29], S M Aronson[30], K G Arun[31], S Ascenzi[16,32], G Ashton[6], S M Aston[7], P Astone[33], F Aubin[34], P Aufmuth[10], K AultONeal[35], C Austin[2], V Avendano[36], A Avila-Alvarez[27], S Babak[26], P Bacon[26], F Badaracco[16,17], M K M Bader[37], S Bae[38], J Baird[26], P T Baker[39], F Baldaccini[40,41], G Ballardin[29], S W Ballmer[42], A Bals[35], S Banagiri[43], J C Barayoga[1], C Barbieri[44,45], S E Barclay[46], B C Barish[1], D Barker[47], K Barkett[48], S Barnum[14], F Barone[5,49], B Barr[46], L Barsotti[14], M Barsuglia[26], D Barta[50], J Bartlett[47], I Bartos[30], R Bassiri[51], A Basti[20,21], M Bawaj[41,52], J C Bayley[46], M Bazzan[53,54], B Bécsy[55], M Bejger[26,56], I Belahcene[28], A S Bell[46], D Beniwal[57], M G Benjamin[35], B K Berger[51], G Bergmann[9,10], S Bernuzzi[11], C P L Berry[58], D Bersanetti[59], A Bertolini[37], J Betzwieser[7], R Bhandare[60], J Bidler[27], E Biggs[24], I A Bilenko[61], S A Bilgili[39], G Billingsley[1], R Birney[25], O Birnholtz[62], S Biscans[1,14], M Bischi[63,64], S Biscoveanu[14], A Bisht[10], M Bitossi[21,29], M A Bizouard[65], J K Blackburn[1], J Blackman[48], C D Blair[7], D G Blair[66], R M Blair[47], S Bloemen[67], F Bobba[68,69], N Bode[9,10], M Boer[65], Y Boetzel[70], G Bogaert[65], F Bondu[71], R Bonnand[34], P Booker[9,10], B A Boom[37], R Bork[1], V Boschi[29], S Bose[3], V Bossilkov[66], J Bosveld[66], Y Bouffanais[53,54], A Bozzi[29], C Bradaschia[21], P R Brady[24], A Bramley[7], M Branchesi[16,17], J E Brau[72], M Breschi[11], T Briant[73], J H Briggs[46], F Brighenti[63,64], A Brillet[65], M Brinkmann[9,10], P Brockill[24],

A F Brooks[1], J Brooks[29], D D Brown[57], S Brunett[1], A Buikema[14], T Bulik[74], H J Bulten[37,75], A Buonanno[76,77], D Buskulic[34], C Buy[26], R L Byer[51], M Cabero[9,10], L Cadonati[78], G Cagnoli[79], C Cahillane[1], J Calderón Bustillo[6], T A Callister[1], E Calloni[5,80], J B Camp[81], W A Campbell[6], M Canepa[59,82], K C Cannon[83], H Cao[57], J Cao[84], G Carapella[68,69], F Carbognani[29], S Caride[85], M F Carney[58], G Carullo[20,21], J Casanueva Diaz[21], C Casentini[32,86], S Caudill[37], M Cavaglià[87,88], F Cavalier[28], R Cavalieri[29], G Cella[21], P Cerdá-Durán[22], E Cesarini[32,89], O Chaibi[65], K Chakravarti[3], S J Chamberlin[90], M Chan[46], S Chao[91], P Charlton[92], E A Chase[58], E Chassande-Mottin[26], D Chatterjee[24], M Chaturvedi[60], K Chatziioannou[93], B D Cheeseboro[39], H Y Chen[94], X Chen[66], Y Chen[48], H-P Cheng[30], C K Cheong[95], H Y Chia[30], F Chiadini[69,96], A Chincarini[59], A Chiummo[29], G Cho[97], H S Cho[98], M Cho[77], N Christensen[65,99], Q Chu[66], S Chua[73], K W Chung[95], S Chung[66], G Ciani[53,54], M Cieślar[56], A A Ciobanu[57], R Ciolfi[54,100], F Cipriano[65], A Cirone[59,82], F Clara[47], J A Clark[78], P Clearwater[101], F Cleva[65], E Coccia[16,17], P-F Cohadon[73], D Cohen[28], M Colleoni[102], C G Collette[103], C Collins[13], M Colpi[44,45], L R Cominsky[104], M Constancio Jr[15], L Conti[54], S J Cooper[13], P Corban[7], T R Corbitt[2], I Cordero-Carrión[105], S Corezzi[40,41], K R Corley[106], N Cornish[55], D Corre[28], A Corsi[85], S Cortese[29], C A Costa[15], R Cotesta[76], M W Coughlin[1], S B Coughlin[58,107], J-P Coulon[65], S T Countryman[106], P Couvares[1], P B Covas[102], E E Cowan[78], D M Coward[66], M J Cowart[7], D C Coyne[1], R Coyne[108], J D E Creighton[24], T D Creighton[109], J Cripe[2], M Croquette[73], S G Crowder[110], T J Cullen[2], A Cumming[46], L Cunningham[46], E Cuoco[29], T Dal Canton[81], G Dálya[111], B D'Angelo[59,82], S L Danilishin[9,10], S D'Antonio[32], K Danzmann[9,10], A Dasgupta[112], C F Da Silva Costa[30], L E H Datrier[46], V Dattilo[29], I Dave[60], M Davier[28], D Davis[42], E J Daw[113], D DeBra[51], M Deenadayalan[3], J Degallaix[23], M De Laurentis[5,80], S Deléglise[73], W Del Pozzo[20,21], L M DeMarchi[58], N Demos[14], T Dent[114], R De Pietri[115,116], R De Rosa[5,80], C De Rossi[23,29], R DeSalvo[117], O de Varona[9,10], S Dhurandhar[3], M C Díaz[109], T Dietrich[37], L Di Fiore[5], C DiFronzo[13], C Di Giorgio[68,69], F Di Giovanni[22], M Di Giovanni[118,119], T Di Girolamo[5,80], A Di Lieto[20,21], B Ding[103], S Di Pace[33,120], I Di Palma[33,120], F Di Renzo[20,21], A K Divakarla[30], A Dmitriev[13], Z Doctor[94], F Donovan[14], K L Dooley[87,107], S Doravari[3], I Dorrington[107], T P Downes[24], M Drago[16,17], J C Driggers[47], Z Du[84], J-G Ducoin[28], P Dupej[46], O Durante[68,69], S E Dwyer[47],

P J Easter[6], G Eddolls[46], T B Edo[113], A Effler[7], P Ehrens[1], J Eichholz[8], S S Eikenberry[30], M Eisenmann[34], R A Eisenstein[14], L Errico[5,80], R C Essick[94], H Estelles[102], D Estevez[34], Z B Etienne[39], T Etzel[1], M Evans[14], T M Evans[7], V Fafone[16,32,86], S Fairhurst[107], X Fan[84], S Farinon[59], B Farr[72], W M Farr[13], E J Fauchon-Jones[107], M Favata[36], M Fays[113], M Fazio[121], C Fee[122], J Feicht[1], M M Fejer[51], F Feng[26], A Fernandez-Galiana[14], I Ferrante[20,21], E C Ferreira[15], T A Ferreira[15], F Fidecaro[20,21], I Fiori[29], D Fiorucci[16,17], M Fishbach[94], R P Fisher[123], J M Fishner[14], R Fittipaldi[69,124], M Fitz-Axen[43], V Fiumara[69,125], R Flaminio[34,126], M Fletcher[46], E Floden[43], E Flynn[27], H Fong[83], J A Font[22,127], P W F Forsyth[8], J-D Fournier[65], S Frasca[33,120], F Frasconi[21], Z Frei[111], A Freise[13], R Frey[72], V Frey[28], P Fritschel[14], V V Frolov[7], G Fronzè[128], P Fulda[30], M Fyffe[7], H A Gabbard[46], B U Gadre[76], S M Gaebel[13], J R Gair[129], L Gammaitoni[40], S G Gaonkar[3], C García-Quirós[102], F Garufi[5,80], B Gateley[47], S Gaudio[35], G Gaur[130], V Gayathri[131], G Gemme[59], E Genin[29], A Gennai[21], D George[19], J George[60], L Gergely[132], S Ghonge[78], Abhirup Ghosh[76], Archisman Ghosh[37], S Ghosh[24], B Giacomazzo[118,119], J A Giaime[2,7], K D Giardina[7], D R Gibson[133], K Gill[106], L Glover[134], J Gniesmer[135], P Godwin[90], E Goetz[47], R Goetz[30], B Goncharov[6], G González[2], J M Gonzalez Castro[20,21], A Gopakumar[136], S E Gossan[1], M Gosselin[20,21,29], R Gouaty[34], B Grace[8], A Grado[5,137], M Granata[23], A Grant[46], S Gras[14], P Grassia[1], C Gray[47], R Gray[46], G Greco[63,64], A C Green[30], R Green[107], E M Gretarsson[35], A Grimaldi[118,119], S J Grimm[16,17], P Groot[67], H Grote[107], S Grunewald[76], P Gruning[28], G M Guidi[63,64], H K Gulati[112], Y Guo[37], A Gupta[90], Anchal Gupta[1], P Gupta[37], E K Gustafson[1], R Gustafson[138], L Haegel[102], O Halim[16,17], B R Hall[139], E D Hall[14], E Z Hamilton[107], G Hammond[46], M Haney[70], M M Hanke[9,10], J Hanks[47], C Hanna[90], M D Hannam[107], O A Hannuksela[95], T J Hansen[35], J Hanson[7], T Harder[65], T Hardwick[2], K Haris[18], J Harms[16,17], G M Harry[140], I W Harry[141], R K Hasskew[7], C J Haster[14], K Haughian[46], F J Hayes[46], J Healy[62], A Heidmann[73], M C Heintze[7], H Heitmann[65], F Hellman[142], P Hello[28], G Hemming[29], M Hendry[46], I S Heng[46], J Hennig[9,10], Francisco Hernandez Vivanco[6], M Heurs[9,10], S Hild[46], T Hinderer[37,143,144], W C G Ho[145], S Hochheim[9,10], D Hofman[23], A M Holgado[19], N A Holland[8], K Holt[7], D E Holz[94], P Hopkins[107], C Horst[24], J Hough[46], E J Howell[66], C G Hoy[107], Y Huang[14], M T Hübner[6], E A Huerta[19], D Huet[28], B Hughey[35], V Hui[34], S Husa[102], S H Huttner[46], T Huynh-Dinh[7], B Idzkowski[74], A Iess[32,86],

H Inchauspe[30], C Ingram[57], R Inta[85], G Intini[33,120], B Irwin[122], H N Isa[46], J-M Isac[73], M Isi[14], B R Iyer[18], T Jacqmin[73], S J Jadhav[146], K Jani[78], N N Janthalur[146], P Jaranowski[147], D Jariwala[30], A C Jenkins[148], J Jiang[30], D S Johnson[19], A W Jones[13], D I Jones[149], J D Jones[47], R Jones[46], R J G Jonker[37], L Ju[66], J Junker[9,10], C V Kalaghatgi[107], V Kalogera[58], B Kamai[1], S Kandhasamy[3], G Kang[38], J B Kanner[1], S J Kapadia[24], S Karki[72], R Kashyap[18], M Kasprzack[1], W Kastaun[9,10], S Katsanevas[29], E Katsavounidis[14], W Katzman[7], S Kaufer[10], K Kawabe[47], N V Keerthana[3], F Kéfélian[65], D Keitel[141], R Kennedy[113], J S Key[150], F Y Khalili[61], I Khan[16,32], S Khan[9,10], E A Khazanov[151], N Khetan[16,17], M Khursheed[60], N Kijbunchoo[8], Chunglee Kim[152], J C Kim[153], K Kim[95], W Kim[57], W S Kim[154], Y-M Kim[155], C Kimball[58], P J King[47], M Kinley-Hanlon[46], R Kirchhoff[9,10], J S Kissel[47], L Kleybolte[135], J H Klika[24], S Klimenko[30], T D Knowles[39], P Koch[9,10], S M Koehlenbeck[9,10], G Koekoek[37,156], S Koley[37], V Kondrashov[1], A Kontos[157], N Koper[9,10], M Korobko[135], W Z Korth[1], M Kovalam[66], D B Kozak[1], C Krämer[9,10], V Kringel[9,10], N Krishnendu[31], A Królak[158,159], N Krupinski[24], G Kuehn[9,10], A Kumar[146], P Kumar[160], Rahul Kumar[47], Rakesh Kumar[112], L Kuo[91], A Kutynia[158], S Kwang[24], B D Lackey[76], D Laghi[20,21], K H Lai[95], T L Lam[95], M Landry[47], P Landry[94], B B Lane[14], R N Lang[161], J Lange[62], B Lantz[51], R K Lanza[14], A Lartaux-Vollard[28], P D Lasky[6], M Laxen[7], A Lazzarini[1], C Lazzaro[54], P Leaci[120,33], S Leavey[9,10], Y K Lecoeuche[47], C H Lee[98], H K Lee[162], H M Lee[163], H W Lee[153], J Lee[97], K Lee[46], J Lehmann[9,10], A K Lenon[39], N Leroy[28], N Letendre[34], Y Levin[6], A Li[95], J Li[84], K J L Li[95], T G F Li[95], X Li[48], F Lin[6], F Linde[37,164], S D Linker[134], T B Littenberg[165], J Liu[66], X Liu[24], M Llorens-Monteagudo[22], R K L Lo[95,1], L T London[14], A Longo[166,167], M Lorenzini[16,17], V Loriette[168], M Lormand[7], G Losurdo[21], J D Lough[9,10], C O Lousto[62], G Lovelace[27], M E Lower[169], H Lück[9,10], D Lumaca[32,86], A P Lundgren[141], R Lynch[14], Y Ma[48], R Macas[107], S Macfoy[25], M MacInnis[14], D M Macleod[107], A Macquet[65], I Magaña Hernandez[24], F Magaña-Sandoval[30], R M Magee[90], E Majorana[33], I Maksimovic[168], A Malik[60], N Man[65], V Mandic[43], V Mangano[33,46,120], G L Mansell[14,47], M Manske[24], M Mantovani[29], M Mapelli[53,54], F Marchesoni[41,52], F Marion[34], S Márka[106], Z Márka[106], C Markakis[19], A S Markosyan[51], A Markowitz[1], E Maros[1], A Marquina[105], S Marsat[26], F Martelli[63,64], I W Martin[46], R M Martin[36], V Martinez[79], D V Martynov[13], H Masalehdan[135], K Mason[14],

E Massera[113], A Masserot[34], T J Massinger[1], M Masso-Reid[46], S Mastrogiovanni[26], A Matas[76], F Matichard[1,14], L Matone[106], N Mavalvala[14], J J McCann[66], R McCarthy[47], D E McClelland[8], S McCormick[7], L McCuller[14], S C McGuire[170], C McIsaac[141], J McIver[1], D J McManus[8], T McRae[8], S T McWilliams[39], D Meacher[24], G D Meadors[6], M Mehmet[9,10], A K Mehta[18], J Meidam[37], E Mejuto Villa[69,117], A Melatos[101], G Mendell[47], R A Mercer[24], L Mereni[23], K Merfeld[72], E L Merilh[47], M Merzougui[65], S Meshkov[1], C Messenger[46], C Messick[90], F Messina[44,45], R Metzdorff[73], P M Meyers[101], F Meylahn[9,10], A Miani[118,119], H Miao[13], C Michel[23], H Middleton[101], L Milano[80,5], A L Miller[30,33,120], M Millhouse[101], J C Mills[107], M C Milovich-Goff[134], O Minazzoli[65,171], Y Minenkov[32], A Mishkin[30], C Mishra[172], T Mistry[113], S Mitra[3], V P Mitrofanov[61], G Mitselmakher[30], R Mittleman[14], G Mo[99], D Moffa[122], K Mogushi[87], S R P Mohapatra[14], M Molina-Ruiz[142], M Mondin[134], M Montani[63,64], C J Moore[13], D Moraru[47], F Morawski[56], G Moreno[47], S Morisaki[83], B Mours[34], C M Mow-Lowry[13], F Muciaccia[33,120], Arunava Mukherjee[9,10], D Mukherjee[24], S Mukherjee[109], Subroto Mukherjee[112], N Mukund[3,9,10], A Mullavey[7], J Munch[57], E A Muñiz[42], M Muratore[35], P G Murray[46], A Nagar[89,128,173], I Nardecchia[32,86], L Naticchioni[33,120], R K Nayak[174], B F Neil[66], J Neilson[69,117], G Nelemans[37,67], T J N Nelson[7], M Nery[9,10], A Neunzert[138], L Nevin[1], K Y Ng[14], S Ng[57], C Nguyen[26], P Nguyen[72], D Nichols[37,143], S A Nichols[2], S Nissanke[37,143], F Nocera[29], C North[107], L K Nuttall[141], M Obergaulinger[22,175], J Oberling[47], B D O'Brien[30], G Oganesyan[16,17], G H Ogin[176], J J Oh[154], S H Oh[154], F Ohme[9,10], H Ohta[83], M A Okada[15], M Oliver[102], P Oppermann[9,10], Richard J Oram[7], B O'Reilly[7], R G Ormiston[43], L F Ortega[30], R O'Shaughnessy[62], S Ossokine[76], D J Ottaway[57], H Overmier[7], B J Owen[85], A E Pace[90], G Pagano[20,21], M A Page[66], G Pagliaroli[16,17], A Pai[131], S A Pai[60], J R Palamos[72], O Palashov[151], C Palomba[33], H Pan[91], P K Panda[146], P T H Pang[37,95], C Pankow[58], F Pannarale[33,120], B C Pant[60], F Paoletti[21], A Paoli[29], A Parida[3], W Parker[7,170], D Pascucci[37,46], A Pasqualetti[29], R Passaquieti[20,21], D Passuello[21], M Patil[159], B Patricelli[20,21], E Payne[6], B L Pearlstone[46], T C Pechsiri[30], A J Pedersen[42], M Pedraza[1], R Pedurand[23,177], A Pele[7], S Penn[178], A Perego[118,119], C J Perez[47], C Périgois[34], A Perreca[118,119], J Petermann[135], H P Pfeiffer[76], M Phelps[9,10], K S Phukon[3], O J Piccinni[33,120], M Pichot[65], F Piergiovanni[63,64], V Pierro[69,117], G Pillant[29], L Pinard[23],

I M Pinto[69,89,117], M Pirello[47], M Pitkin[46], W Plastino[166,167],
R Poggiani[20,21], D Y T Pong[95], S Ponrathnam[3], P Popolizio[29],
E K Porter[26], J Powell[169], A K Prajapati[112], J Prasad[3],
K Prasai[51], R Prasanna[146], G Pratten[102], T Prestegard[24],
M Principe[89,69,117], G A Prodi[118,119], L Prokhorov[13],
M Punturo[41], P Puppo[33], M Pürrer[76], H Qi[107], V Quetschke[109],
P J Quinonez[35], F J Raab[47], G Raaijmakers[37,143], H Radkins[47],
N Radulesco[65], P Raffai[111], S Raja[60], C Rajan[60],
B Rajbhandari[85], M Rakhmanov[109], K E Ramirez[109],
A Ramos-Buades[102], Javed Rana[3], K Rao[58],
P Rapagnani[33,120], V Raymond[107], M Razzano[20,21], J Read[27],
T Regimbau[34], L Rei[59], S Reid[25], D H Reitze[1,30],
P Rettegno[128,179], F Ricci[120,33], C J Richardson[35],
J W Richardson[1], P M Ricker[19], G Riemenschneider[128,179],
K Riles[138], M Rizzo[58], N A Robertson[1,46], F Robinet[28],
A Rocchi[32], L Rolland[34], J G Rollins[1], V J Roma[72],
M Romanelli[71], R Romano[4,5], C L Romel[47], J H Romie[7],
C A Rose[24], D Rose[27], K Rose[122], D Rosińska[74],
S G Rosofsky[19], M P Ross[180], S Rowan[46], A Rüdiger[9,10],
P Ruggi[29], G Rutins[133], K Ryan[47], S Sachdev[90], T Sadecki[47],
M Sakellariadou[148], O S Salafia[44,45,181], L Salconi[29],
M Saleem[31], A Samajdar[37], L Sammut[6], E J Sanchez[1],
L E Sanchez[1], N Sanchis-Gual[182], J R Sanders[183],
K A Santiago[36], E Santos[65], N Sarin[6], B Sassolas[23],
B S Sathyaprakash[90,107], O Sauter[138,34], R L Savage[47],
P Schale[72], M Scheel[48], J Scheuer[58], P Schmidt[13,67],
R Schnabel[135], R M S Schofield[72], A Schönbeck[135],
E Schreiber[9,10], B W Schulte[9,10], B F Schutz[107], J Scott[46],
S M Scott[8], E Seidel[19], D Sellers[7], A S Sengupta[184],
N Sennett[76], D Sentenac[29], V Sequino[59], A Sergeev[151],
Y Setyawati[9,10], D A Shaddock[8], T Shaffer[47], M S Shahriar[58],
M B Shaner[134], A Sharma[16,17], P Sharma[60], P Shawhan[77],
H Shen[19], R Shink[185], D H Shoemaker[14], D M Shoemaker[78],
K Shukla[142], S ShyamSundar[60], K Siellez[78], M Sieniawska[56],
D Sigg[47], L P Singer[81], D Singh[90], N Singh[74], A Singhal[16,33],
A M Sintes[102], S Sitmukhambetov[109], V Skliris[107],
B J J Slagmolen[8], T J Slaven-Blair[66], J R Smith[27],
R J E Smith[6], S Somala[186], E J Son[154], S Soni[2], B Sorazu[46],
F Sorrentino[59], T Souradeep[3], E Sowell[85], A P Spencer[46],
M Spera[53,54], A K Srivastava[112], V Srivastava[42], K Staats[58],
C Stachie[65], M Standke[9,10], D A Steer[26], M Steinke[9,10],
J Steinlechner[46,135], S Steinlechner[135], D Steinmeyer[9,10],
S P Stevenson[169], D Stocks[51], R Stone[109], D J Stops[13],
K A Strain[46], G Stratta[64,187], S E Strigin[61], A Strunk[47],
R Sturani[188], A L Stuver[189], V Sudhir[14], T Z Summerscales[190],

**L Sun[1], S Sunil[112], A Sur[56], J Suresh[83], P J Sutton[107], B L Swinkels[37], M J Szczepańczyk[35], M Tacca[37], S C Tait[46], C Talbot[6], D B Tanner[30], D Tao[1], M Tápai[132], A Tapia[27], J D Tasson[99], R Taylor[1], R Tenorio[102], L Terkowski[135], M Thomas[7], P Thomas[47], S R Thondapu[60], K A Thorne[7], E Thrane[6], Shubhanshu Tiwari[118,119], Srishti Tiwari[136], V Tiwari[107], K Toland[46], M Tonelli[20,21], Z Tornasi[46], A Torres-Forné[191], C I Torrie[1], D Töyrä[13], F Travasso[29,41], G Traylor[7], M C Tringali[74], A Tripathee[138], A Trovato[26], L Trozzo[21,192], K W Tsang[37], M Tse[14], R Tso[48], L Tsukada[83], D Tsuna[83], T Tsutsui[83], D Tuyenbayev[109], K Ueno[83], D Ugolini[193], C S Unnikrishnan[136], A L Urban[2], S A Usman[94], H Vahlbruch[10], G Vajente[1], G Valdes[2], M Valentini[118,119], N van Bakel[37], M van Beuzekom[37], J F J van den Brand[37,75], C Van Den Broeck[37,194], D C Vander-Hyde[42], L van der Schaaf[37], J V VanHeijningen[66], A A van Veggel[46], M Vardaro[53,54], V Varma[48], S Vass[1], M Vasúth[50], A Vecchio[13], G Vedovato[54], J Veitch[46], P J Veitch[57], K Venkateswara[180], G Venugopalan[1], D Verkindt[34], F Vetrano[63,64], A Viceré[63,64], A D Viets[24], S Vinciguerra[13], D J Vine[133], J-Y Vinet[65], S Vitale[14], T Vo[42], H Vocca[40,41], C Vorvick[47], S P Vyatchanin[61], A R Wade[1], L E Wade[122], M Wade[122], R Walet[37], M Walker[27], L Wallace[1], S Walsh[24], H Wang[13], J Z Wang[138], S Wang[19], W H Wang[109], Y F Wang[95], R L Ward[8], Z A Warden[35], J Warner[47], M Was[34], J Watchi[103], B Weaver[47], L-W Wei[9,10], M Weinert[9,10], A J Weinstein[1], R Weiss[14], F Wellmann[9,10], L Wen[66], E K Wessel[19], P Weßels[9,10], J W Westhouse[35], K Wette[8], J T Whelan[62], B F Whiting[30], C Whittle[14], D M Wilken[9,10], D Williams[46], A R Williamson[37,143], J L Willis[1], B Willke[10,9], W Winkler[9,10], C C Wipf[1], H Wittel[9,10], G Woan[46], J Woehler[9,10], J K Wofford[62], J L Wright[46], D S Wu[9,10], D M Wysocki[62], S Xiao[1], R Xu[110], H Yamamoto[1], C C Yancey[77], L Yang[121], Y Yang[30], Z Yang[43], M J Yap[8], M Yazback[30], D W Yeeles[107], Hang Yu[14], Haocun Yu[14], S H R Yuen[95], A K Zadrożny[109], A Zadrożny[158], M Zanolin[35], F Zappa[76], T Zelenova[29], J-P Zendri[54], M Zevin[58], J Zhang[66], L Zhang[1], T Zhang[46], C Zhao[66], G Zhao[103], M Zhou[58], Z Zhou[58], X J Zhu[6], A B Zimmerman[195], Y Zlochower[62], M E Zucker[1,14], J Zweizig[1]**
**(The LIGO Scientific Collaboration and The Virgo Collaboration)**

[1] LIGO, California Institute of Technology, Pasadena, CA 91125, United States of America
[2] Louisiana State University, Baton Rouge, LA 70803, United States of America
[3] Inter-University Centre for Astronomy and Astrophysics, Pune 411007, India
[4] Dipartimento di Farmacia, Università di Salerno, I-84084 Fisciano, Salerno, Italy

[5] INFN, Sezione di Napoli, Complesso Universitario di Monte S.Angelo, I-80126 Napoli, Italy
[6] OzGrav, School of Physics & Astronomy, Monash University, Clayton 3800, Victoria, Australia
[7] LIGO Livingston Observatory, Livingston, LA 70754, United States of America
[8] OzGrav, Australian National University, Canberra, Australian Capital Territory 0200, Australia
[9] Max Planck Institute for Gravitational Physics (Albert Einstein Institute), D-30167 Hannover, Germany
[10] Leibniz Universität Hannover, D-30167 Hannover, Germany
[11] Theoretisch-Physikalisches Institut, Friedrich-Schiller-Universität Jena, D-07743 Jena, Germany
[12] University of Cambridge, Cambridge CB2 1TN, United Kingdom
[13] University of Birmingham, Birmingham B15 2TT, United Kingdom
[14] LIGO, Massachusetts Institute of Technology, Cambridge, MA 02139, United States of America
[15] Instituto Nacional de Pesquisas Espaciais, 12227-010 São José dos Campos, São Paulo, Brazil
[16] Gran Sasso Science Institute (GSSI), I-67100 L'Aquila, Italy
[17] INFN, Laboratori Nazionali del Gran Sasso, I-67100 Assergi, Italy
[18] International Centre for Theoretical Sciences, Tata Institute of Fundamental Research, Bengaluru 560089, India
[19] NCSA, University of Illinois at Urbana-Champaign, Urbana, IL 61801, United States of America
[20] Università di Pisa, I-56127 Pisa, Italy
[21] INFN, Sezione di Pisa, I-56127 Pisa, Italy
[22] Departamento de Astronomía y Astrofísica, Universitat de València, E-46100 Burjassot, València, Spain
[23] Laboratoire des Matériaux Avancés (LMA), CNRS/IN2P3, F-69622 Villeurbanne, France
[24] University of Wisconsin-Milwaukee, Milwaukee, WI 53201, United States of America
[25] SUPA, University of Strathclyde, Glasgow G1 1XQ, United Kingdom
[26] APC, AstroParticule et Cosmologie, Université Paris Diderot, CNRS/IN2P3, CEA/Irfu, Observatoire de Paris, Sorbonne Paris Cité, F-75205 Paris Cedex 13, France
[27] California State University Fullerton, Fullerton, CA 92831, United States of America
[28] LAL, Univ. Paris-Sud, CNRS/IN2P3, Université Paris-Saclay, F-91898 Orsay, France
[29] European Gravitational Observatory (EGO), I-56021 Cascina, Pisa, Italy
[30] University of Florida, Gainesville, FL 32611, United States of America
[31] Chennai Mathematical Institute, Chennai 603103, India
[32] INFN, Sezione di Roma Tor Vergata, I-00133 Roma, Italy
[33] INFN, Sezione di Roma, I-00185 Roma, Italy
[34] Laboratoire d'Annecy de Physique des Particules (LAPP), Univ. Grenoble Alpes, Université Savoie Mont Blanc, CNRS/IN2P3, F-74941 Annecy, France
[35] Embry-Riddle Aeronautical University, Prescott, AZ 86301, United States of America
[36] Montclair State University, Montclair, NJ 07043, United States of America
[37] Nikhef, Science Park 105, 1098 XG Amsterdam, The Netherlands
[38] Korea Institute of Science and Technology Information, Daejeon 34141, Republic of Korea
[39] West Virginia University, Morgantown, WV 26506, United States of America

[40] Università di Perugia, I-06123 Perugia, Italy
[41] INFN, Sezione di Perugia, I-06123 Perugia, Italy
[42] Syracuse University, Syracuse, NY 13244, United States of America
[43] University of Minnesota, Minneapolis, MN 55455, United States of America
[44] Università degli Studi di Milano-Bicocca, I-20126 Milano, Italy
[45] INFN, Sezione di Milano-Bicocca, I-20126 Milano, Italy
[46] SUPA, University of Glasgow, Glasgow G12 8QQ, United Kingdom
[47] LIGO Hanford Observatory, Richland, WA 99352, United States of America
[48] Caltech CaRT, Pasadena, CA 91125, United States of America
[49] Dipartimento di Medicina, Chirurgia e Odontoiatria 'Scuola Medica Salernitana', Università di Salerno, I-84081 Baronissi, Salerno, Italy
[50] Wigner RCP, RMKI, H-1121 Budapest, Konkoly Thege Miklós út 29-33, Hungary
[51] Stanford University, Stanford, CA 94305, United States of America
[52] Università di Camerino, Dipartimento di Fisica, I-62032 Camerino, Italy
[53] Università di Padova, Dipartimento di Fisica e Astronomia, I-35131 Padova, Italy
[54] INFN, Sezione di Padova, I-35131 Padova, Italy
[55] Montana State University, Bozeman, MT 59717, United States of America
[56] Nicolaus Copernicus Astronomical Center, Polish Academy of Sciences, 00-716, Warsaw, Poland
[57] OzGrav, University of Adelaide, Adelaide, South Australia 5005, Australia
[58] Center for Interdisciplinary Exploration & Research in Astrophysics (CIERA), Northwestern University, Evanston, IL 60208, United States of America
[59] INFN, Sezione di Genova, I-16146 Genova, Italy
[60] RRCAT, Indore, Madhya Pradesh 452013, India
[61] Faculty of Physics, Lomonosov Moscow State University, Moscow 119991, Russia
[62] Rochester Institute of Technology, Rochester, NY 14623, United States of America
[63] Università degli Studi di Urbino 'Carlo Bo', I-61029 Urbino, Italy
[64] INFN, Sezione di Firenze, I-50019 Sesto Fiorentino, Firenze, Italy
[65] Artemis, Université Côte d'Azur, Observatoire Côte d'Azur, CNRS, CS 34229, F-06304 Nice Cedex 4, France
[66] OzGrav, University of Western Australia, Crawley, Western Australia 6009, Australia
[67] Department of Astrophysics/IMAPP, Radboud University Nijmegen, PO Box 9010, 6500 GL Nijmegen, The Netherlands
[68] Dipartimento di Fisica 'E.R. Caianiello', Università di Salerno, I-84084 Fisciano, Salerno, Italy
[69] INFN, Sezione di Napoli, Gruppo Collegato di Salerno, Complesso Universitario di Monte S. Angelo, I-80126 Napoli, Italy
[70] Physik-Institut, University of Zurich, Winterthurerstrasse 190, 8057 Zurich, Switzerland
[71] Univ Rennes, CNRS, Institut FOTON—UMR6082, F-3500 Rennes, France
[72] University of Oregon, Eugene, OR 97403, United States of America
[73] Laboratoire Kastler Brossel, Sorbonne Université, CNRS, ENS-Université PSL, Collège de France, F-75005 Paris, France
[74] Astronomical Observatory Warsaw University, 00-478 Warsaw, Poland
[75] VU University Amsterdam, 1081 HV Amsterdam, The Netherlands
[76] Max Planck Institute for Gravitational Physics (Albert Einstein Institute), D-14476 Potsdam-Golm, Germany
[77] University of Maryland, College Park, MD 20742, United States of America
[78] School of Physics, Georgia Institute of Technology, Atlanta, GA 30332, United States of America
[79] Université de Lyon, Université Claude Bernard Lyon 1, CNRS, Institut Lumière Matière, F-69622 Villeurbanne, France

[80] Università di Napoli 'Federico II', Complesso Universitario di Monte S.Angelo, I-80126 Napoli, Italy
[81] NASA Goddard Space Flight Center, Greenbelt, MD 20771, United States of America
[82] Dipartimento di Fisica, Università degli Studi di Genova, I-16146 Genova, Italy
[83] RESCEU, University of Tokyo, Tokyo, 113-0033, Japan
[84] Tsinghua University, Beijing 100084, People's Republic of China
[85] Texas Tech University, Lubbock, TX 79409, United States of America
[86] Università di Roma Tor Vergata, I-00133 Roma, Italy
[87] The University of Mississippi, University, MS 38677, United States of America
[88] Missouri University of Science and Technology, Rolla, MO 65409, United States of America
[89] Museo Storico della Fisica e Centro Studi e Ricerche 'Enrico Fermi', I-00184 Roma, Italy
[90] The Pennsylvania State University, University Park, PA 16802, United States of America
[91] National Tsing Hua University, Hsinchu City, 30013 Taiwan, Republic of China
[92] Charles Sturt University, Wagga Wagga, New South Wales 2678, Australia
[93] Canadian Institute for Theoretical Astrophysics, University of Toronto, Toronto, Ontario M5S 3H8, Canada
[94] University of Chicago, Chicago, IL 60637, United States of America
[95] The Chinese University of Hong Kong, Shatin, NT, Hong Kong
[96] Dipartimento di Ingegneria Industriale (DIIN), Università di Salerno, I-84084 Fisciano, Salerno, Italy
[97] Seoul National University Seoul 08826, Republic of Korea
[98] PUnited States of American National University, Busan 46241, Republic of Korea
[99] Carleton College, Northfield, MN 55057, United States of America
[100] INAF, Osservatorio Astronomico di Padova, I-35122 Padova, Italy
[101] OzGrav, University of Melbourne, Parkville, Victoria 3010, Australia
[102] Universitat de les Illes Balears, IAC3—IEEC, E-07122 Palma de Mallorca, Spain
[103] Université Libre de Bruxelles, Brussels 1050, Belgium
[104] Sonoma State University, Rohnert Park, CA 94928, United States of America
[105] Departamento de Matemáticas, Universitat de València, E-46100 Burjassot, València, Spain
[106] Columbia University, New York, NY 10027, United States of America
[107] Cardiff University, Cardiff CF24 3AA, United Kingdom
[108] University of Rhode Island, Kingston, RI 02881, United States of America
[109] The University of Texas Rio Grande Valley, Brownsville, TX 78520, United States of America
[110] Bellevue College, Bellevue, WA 98007, United States of America
[111] MTA-ELTE Astrophysics Research Group, Institute of Physics, Eötvös University, Budapest 1117, Hungary
[112] Institute for Plasma Research, Bhat, Gandhinagar 382428, India
[113] The University of Sheffield, Sheffield S10 2TN, United Kingdom
[114] IGFAE, Campus Sur, Universidade de Santiago de Compostela, 15782 Spain
[115] Dipartimento di Scienze Matematiche, Fisiche e Informatiche, Università di Parma, I-43124 Parma, Italy
[116] INFN, Sezione di Milano Bicocca, Gruppo Collegato di Parma, I-43124 Parma, Italy
[117] Dipartimento di Ingegneria, Università del Sannio, I-82100 Benevento, Italy
[118] Dipartimento di Fisica, Università di Trento, I-38123 Povo, Trento, Italy
[119] INFN, Trento Institute for Fundamental Physics and Applications, I-38123 Povo, Trento, Italy
[120] Università di Roma 'La Sapienza',' I-00185 Roma, Italy

[121] Colorado State University, Fort Collins, CO 80523, United States of America
[122] Kenyon College, Gambier, OH 43022, United States of America
[123] Christopher Newport University, Newport News, VA 23606, United States of America
[124] CNR-SPIN, c/o Università di Salerno, I-84084 Fisciano, Salerno, Italy
[125] Scuola di Ingegneria, Università della Basilicata, I-85100 Potenza, Italy
[126] National Astronomical Observatory of Japan, 2-21-1 Osawa, Mitaka, Tokyo 181-8588, Japan
[127] Observatori Astronòmic, Universitat de València, E-46980 Paterna, València, Spain
[128] INFN Sezione di Torino, I-10125 Torino, Italy
[129] School of Mathematics, University of Edinburgh, Edinburgh EH9 3FD, United Kingdom
[130] Institute Of Advanced Research, Gandhinagar 382426, India
[131] Indian Institute of Technology Bombay, Powai, Mumbai 400 076, India
[132] University of Szeged, Dóm tér 9, Szeged 6720, Hungary
[133] SUPA, University of the West of Scotland, Paisley PA1 2BE, United Kingdom
[134] California State University, Los Angeles, 5151 State University Dr, Los Angeles, CA 90032, United States of America
[135] Universität Hamburg, D-22761 Hamburg, Germany
[136] Tata Institute of Fundamental Research, Mumbai 400005, India
[137] INAF, Osservatorio Astronomico di Capodimonte, I-80131 Napoli, Italy
[138] University of Michigan, Ann Arbor, MI 48109, United States of America
[139] Washington State University, Pullman, WA 99164, United States of America
[140] American University, Washington, D.C. 20016, United States of America
[141] University of Portsmouth, Portsmouth, PO1 3FX, United Kingdom
[142] University of California, Berkeley, CA 94720, United States of America
[143] GRAPPA, Anton Pannekoek Institute for Astronomy and Institute for High-Energy Physics, University of Amsterdam, Science Park 904, 1098 XH Amsterdam, The Netherlands
[144] Delta Institute for Theoretical Physics, Science Park 904, 1090 GL Amsterdam, The Netherlands
[145] Department of Physics and Astronomy, Haverford College, Haverford, PA 19041, United States of America
[146] Directorate of Construction, Services & Estate Management, Mumbai 400094 India
[147] University of Białystok, 15-424 Białystok, Poland
[148] King's College London, University of London, London WC2R 2LS, United Kingdom
[149] University of Southampton, Southampton SO17 1BJ, United Kingdom
[150] University of Washington Bothell, Bothell, WA 98011, United States of America
[151] Institute of Applied Physics, Nizhny Novgorod, 603950, Russia
[152] Ewha Womans University, Seoul 03760, Republic of Korea
[153] Inje University Gimhae, South Gyeongsang 50834, Republic of Korea
[154] National Institute for Mathematical Sciences, Daejeon 34047, Republic of Korea
[155] Ulsan National Institute of Science and Technology, Ulsan 44919, Republic of Korea
[156] Maastricht University, PO Box 616, 6200 MD Maastricht, The Netherlands
[157] Bard College, 30 Campus Rd, Annandale-On-Hudson, NY 12504, United States of America
[158] NCBJ, 05-400 Świerk-Otwock, Poland
[159] Institute of Mathematics, Polish Academy of Sciences, 00656 Warsaw, Poland
[160] Cornell University, Ithaca, NY 14850, United States of America

[161] Hillsdale College, Hillsdale, MI 49242, United States of America
[162] Hanyang University, Seoul 04763, Republic of Korea
[163] Korea Astronomy and Space Science Institute, Daejeon 34055, Republic of Korea
[164] Institute for High-Energy Physics, University of Amsterdam, Science Park 904, 1098 XH Amsterdam, The Netherlands
[165] NASA Marshall Space Flight Center, Huntsville, AL 35811, United States of America
[166] Dipartimento di Matematica e Fisica, Università degli Studi Roma Tre, I-00146 Roma, Italy
[167] INFN, Sezione di Roma Tre, I-00146 Roma, Italy
[168] ESPCI, CNRS, F-75005 Paris, France
[169] OzGrav, Swinburne University of Technology, Hawthorn VIC 3122, Australia
[170] Southern University and A&M College, Baton Rouge, LA 70813, United States of America
[171] Centre Scientifique de Monaco, 8 quai Antoine Ier, MC-98000, Monaco
[172] Indian Institute of Technology Madras, Chennai 600036, India
[173] Institut des Hautes Etudes Scientifiques, F-91440 Bures-sur-Yvette, France
[174] IISER-Kolkata, Mohanpur, West Bengal 741252, India
[175] Institut für Kernphysik, Theoriezentrum, 64289 Darmstadt, Germany
[176] Whitman College, 345 Boyer Avenue, Walla Walla, WA 99362 United States of America
[177] Université de Lyon, F-69361 Lyon, France
[178] Hobart and William Smith Colleges, Geneva, NY 14456, United States of America
[179] Dipartimento di Fisica, Università degli Studi di Torino, I-10125 Torino, Italy
[180] University of Washington, Seattle, WA 98195, United States of America
[181] INAF, Osservatorio Astronomico di Brera sede di Merate, I-23807 Merate, Lecco, Italy
[182] Centro de Astrofísica e Gravitação (CENTRA), Departamento de Física, Instituto Superior Técnico, Universidade de Lisboa, 1049-001 Lisboa, Portugal
[183] Marquette University, 11420W. Clybourn St., Milwaukee, WI 53233, United States of America
[184] Indian Institute of Technology, Gandhinagar Ahmedabad Gujarat 382424, India
[185] Université de Montréal/Polytechnique, Montreal, Quebec H3T 1J4, Canada
[186] Indian Institute of Technology Hyderabad, Sangareddy, Khandi, Telangana 502285, India
[187] INAF, Osservatorio di Astrofisica e Scienza dello Spazio, I-40129 Bologna, Italy
[188] International Institute of Physics, Universidade Federal do Rio Grande do Norte, Natal RN 59078-970, Brazil
[189] Villanova University, 800 Lancaster Ave, Villanova, PA 19085, United States of America
[190] Andrews University, Berrien Springs, MI 49104, United States of America
[191] Max Planck Institute for Gravitationalphysik (Albert Einstein Institute), D-14476 Potsdam-Golm, Germany
[192] Università di Siena, I-53100 Siena, Italy
[193] Trinity University, San Antonio, TX 78212, United States of America
[194] Van Swinderen Institute for Particle Physics and Gravity, University of Groningen, Nijenborgh 4, 9747 AG Groningen, The Netherlands
[195] Department of Physics, University of Texas, Austin, TX 78712, United States of America

E-mail: lsc-spokesperson@ligo.org and virgo-spokesperson@ego-gw.it

CrossMark

**Abstract**

GW170817 is the very first observation of gravitational waves originating from the coalescence of two compact objects in the mass range of neutron stars, accompanied by electromagnetic counterparts, and offers an opportunity to directly probe the internal structure of neutron stars. We perform Bayesian model selection on a wide range of theoretical predictions for the neutron star equation of state. For the binary neutron star hypothesis, we find that we cannot rule out the majority of theoretical models considered. In addition, the gravitational-wave data alone does not rule out the possibility that one or both objects were low-mass black holes. We discuss the possible outcomes in the case of a binary neutron star merger, finding that all scenarios from prompt collapse to long-lived or even stable remnants are possible. For long-lived remnants, we place an upper limit of 1.9 kHz on the rotation rate. If a black hole was formed any time after merger and the coalescing stars were slowly rotating, then the maximum baryonic mass of non-rotating neutron stars is at most $3.05\,M_{\odot}$, and three equations of state considered here can be ruled out. We obtain a tighter limit of $2.67\,M_{\odot}$ for the case that the merger results in a hypermassive neutron star.

Keywords: neutron stars, neutron star equation of state, gravitational wave astronomy, compact object mergers

(Some figures may appear in colour only in the online journal)

## 1. Introduction

On August 17, 2017, the Advanced LIGO [1] and Advanced Virgo [2] observatories detected a gravitational-wave signal, GW170817, from a low-mass compact binary system [3]. The signal was coincident with the gamma-ray burst GRB 170817A [4] and was followed by other observations spanning the electromagnetic spectrum [5]. These electromagnetic observations, coupled with measurements of the component masses from the gravitational-wave observation [6, 7], provide compelling evidence that GW170817 was produced from the merger of two neutron stars.

One of the key goals of observing such neutron star mergers is to extract information regarding the internal structure of neutron stars and to improve our understanding of their highly uncertain equation of state [8, 9]. Information regarding the equation of state is encoded in the variation of the orbital evolution induced by tidal interactions and the deformation of the neutron stars under their own angular momenta [10–14]. Such interactions are small at low frequencies but rapidly increase in strength in the final tens of gravitational-wave cycles before merger. Additional information about the equation of state is carried in the post-merger signal, as the remnant body can be a black hole, a stable neutron star or a meta-stable neutron star that collapses after some delay [15–21]. However, gravitational waves from the post-merger remnant are expected at higher frequencies where the detectors are less sensitive. No such post-merger signal was observed for GW170817 [6, 20, 22, 23].

The difficulty of computing the equation of state from first principles has led to a wide range of plausible theoretical models based on nuclear theoretical calculations. Astronomical observations of pulsars and x-ray binaries (e.g. [8, 24, 25]) have already been able to rule out a subset of proposed models, most notably by measurements of large neutron star masses [26]. Several studies have derived constraints on the equation of state through direct measurements of the neutron star mass and tidal deformability as inferred from GW170817 [27–41].

In this paper, we address whether we can rule out specific equations of state from the gravitational-wave data alone and whether we can rule out the possibility that one or both of the coalescing objects was a low-mass black hole. We opt to work with a subset of equations of state representing nuclear matter properties that vary over a wide range of possible values. The correlations between the properties of the neutron star with the underlying equation of state manifests itself in a wide range of macroscopic physical observables, such as the mass, radius or tidal deformability. In practice, we do not expect the true equation of state to be contained in any of the models used in this analysis. By comparing the Bayesian odds ratio for different models, we expect that equations of state that are most similar to the real equation of state will lead to the highest odds ratios whereas equations of state that differ significantly from the data will be down-ranked in the analysis, giving insight into the physical properties of the true equation of state.

We apply Bayesian techniques [42, 43] to compare a large set of equations of state against an alternative hypothesis. The first fully Bayesian study of the feasibility of probing the neutron star equation of state was presented in [44] and expanded upon in [45]. We consider two possible baseline hypotheses: the first hypothesis is that both component objects were low-mass black holes; the second hypothesis is that GW170817 was composed of two neutron stars with an equation of state corresponding to one of the best fit equations of state in our analysis. In order to make quantitative statements about the relative likelihoods of different equations of state, we do not allow the tidal parameters to vary independently and instead enforce that each neutron star obeys a specific equation of state, similar to [38, 39, 44–46]. This is in direct contrast to the analysis presented in [6], where the tidal parameters were allowed to vary independently implicitly allowing each neutron star to have a different equation of state. By performing parameter estimation using a fixed equation of state, we are able to conveniently incorporate a number of physical prior constraints, as was discussed in detail in [39].

We find that the equations of state yielding excessively large tidal deformabilities are disfavored by the data. For example, the MS1_PP model [47] is disfavored with a Bayes factor that is less than $\sim$0.005 compared to the most favored equation of state models. However, we find that we cannot comprehensively rule out the majority of equations of state used in this analysis. Instead, we find that many of the equations of state have comparable evidences that are within an order of magnitude. This includes the models with the lowest tidal deformabilities as well as the binary black hole case.

In a previous analysis of GW170817 [39], we computed radii and tidal deformabilities for GW170817 using two methods. The first method links the tidal deformabilities of the two neutron stars by assuming that they obey the same equation of state [48, 49] and the second method utilized a parametrized equation of state [50, 51]. The neutron star radii were found to be no larger than $\sim$13 km, disfavoring large radii and stiffer equations of state. In this paper, we perform a direct model comparison between equation of state models, finding results that are consistent with [39]. The equation of state models that are most preferred by the data in this analysis are the ones that predict radii close to those computed in [39].

In addition, we also explore the inferences that can be made about the fate of the remnant object for each proposed equation of state. The range of maximum neutron star masses for the equation of state models allows remnants which are indefinitely stable neutron stars, long-lived supramassive neutron stars, or short-lived hypermassive neutron stars. Prompt collapse remains a possibility, though it can be ruled out for many models. In other cases, the required mass threshold from numerical relativity is not available or is insufficiently accurate to make a clear statement. For cases resulting in a supramassive neutron star, we use the mass range inferred for each of our equation of state models to derive the maximum rotation rate of the neutron star after it settles into a uniformly rotating state, obtaining an upper limit of 1.9 kHz.

Finally, we investigate constraints for the maximum mass of non-rotating neutron stars that follow from different assumptions about the type of the remnant. This is motivated by one of the possible explanations for the short gamma-ray burst models, which requires the formation of a black hole surrounded by a disk [52]. The other model assumes a rapidly rotating magnetar instead [53]. Which one is the correct description for the case at hand is an open question, but if we do assume the first one, then the presence of a black hole implies that neutron stars with the total baryonic mass inferred for the initial binary cannot be stable. The resulting limit is not very restrictive, but would rule out the three most extreme of our equations of state when assuming that the coalescing stars are slowly rotating. The bound can be further tightened by limiting the lifetime of the remnant to the observed delay of the short gamma-ray burst. The steep gradient of lifetime as function of mass near the maximum rotating neutron star mass then places the remnant mass in (or at least close to) the hypermassive range. Our limits are compatible with similar calculations carried out in [27, 29, 54], but are obtained using the fully consistent probability distributions computed for each equation of state model.

## 2. Methodology

### 2.1. Gravitational-wave Bayesian model comparison

In this work our goal is to perform a direct comparison of theoretical models for the neutron star equation of state against the gravitational-wave data for GW170817. In order to do so, we employ Bayesian model selection to test a specific equation of state against a competing hypothesis. We adopt both the low-mass binary black hole hypothesis and a best fit equation of state as baseline hypotheses and assess the impact this has on the inferences that can be made from the data. Detailed introductions to Bayesian methods in gravitational wave astronomy can be found in [42, 55–57].

Given the observed gravitational-wave data $s$ and any prior information $I$, the posterior odds in favor of a given model hypothesis $M_i$ can be expressed using the *odds ratio*

$$O_{12} = \frac{P(M_1|s,I)}{P(M_2|s,I)} = \frac{P(s|M_1,I)}{P(s|M_2,I)}\frac{P(M_1|I)}{P(M_2|I)} = B_{12}\frac{P(M_1|I)}{P(M_2|I)}, \quad (1)$$

where $\mathcal{Z} = P(s|M_i,I)$ is the marginalized likelihood or *evidence* and $P(M_i|I)$ is the prior probability on the model hypothesis. If we assume that the two models are *a priori* equally likely, $P(M_1|I) = P(M_2|I)$, then the odds ratio $O_{12}$ reduces to the Bayes factor $B_{12}$. The evidence is calculated by marginalizing over all parameters weighted by their *prior probabilities* $P(\xi|M_i,I)$

$$\mathcal{Z} = P(s|M_i,I) \propto \int P(s|\xi,M_i,I)\,P(\xi|M_i,I)\,\mathrm{d}\xi, \quad (2)$$

where the likelihood of obtaining the data realization $s$ given a gravitational-wave signal with physical parameters $\xi$ is

$$P(s|\xi, M_i, I) \propto \mathrm{e}^{-\langle s-h(t,\xi) \,|\, s-h(t,\xi) \rangle/2}. \quad (3)$$

Together with a given set of prior probabilities, we have all the information needed to evaluate the evidence in equation (2). In the following subsections, we discuss the waveform models used to describe the gravitational-wave signal $h(t)$ and their dependence on the underlying physical parameters $\xi$.

### 2.2. Prior probabilities

An important consideration when computing the Bayes factors is the choice of prior probabilities. This choice will enter the analysis in two distinct ways. First, we must make a choice for the relative prior probabilities between any two models; $P(M_i|I)$. A wide range of equations of state are still consistent with observations [58], though additional theoretical considerations and observations could allow us to restrict the range of plausible equations of state [3, 6, 59]. Here, neglecting such additional information, we assume that the prior probability for all equations of state is the same. As detailed above, the odds ratio then reduces to the Bayes factor between any two models. The second choice of priors that we must make is the choice of prior probabilities on the parameters that enter a specific model, $P(\xi|M_i, I)$. As the choice of priors will affect the resulting Bayes factors, we explicitly detail the choices made in our analysis below.

The choice of priors that we use in this work is broadly similar to our previous analysis of GW170817 [3, 6]. We assume *a priori* that the source of GW170817 resides in NGC 4993, allowing us to adopt a fixed sky location. As in [6], although the distance to NGC 4993 is known, we infer the luminosity distance purely from the gravitational-wave data. The distance prior assumes that sources are uniformly distributed in volume up to a maximum distance of 75 Mpc. The coalescence time $t_c$ and phase $\phi_c$ are taken to be uniformly distributed and we assume that the binary orientation and polarization angle are isotropic.

The remaining parameters that we sample are the masses and spins of the two compact objects. The choice of prior ranges for these parameters is not so easily made. Observations of neutron stars in our galaxy can give us indications of the mass and spin distributions of binary neutron star systems [8, 60–66], but there is some uncertainty in these measurements. Additionally, these observations would not be appropriate to use if we want to assess the possibility that the system might be two black holes.

For these reasons we consider two sets of priors in this work. The first *narrow* prior is motivated by observed binary neutron star systems in our galaxy [26, 67]. For this prior, we assume that the prior on the component masses is Gaussian with a mean of $1.33 M_\odot$ and a standard deviation of $0.09 M_\odot$ [8]. The prior on the dimensionless spin magnitudes are taken to be uniformly distributed between 0 and 0.05 and the spin directions are assumed to be isotropic. This spin prior is consistent with the observed population of binary neutron stars that will merge within a Hubble time [3, 68]. As many of the waveform models we consider do not allow for spin components perpendicular to the orbital angular momentum, we choose to sample the spin direction and magnitude as described above and then ignore any spin component perpendicular to the orbital angular momentum. In practice, this amounts to choosing a non-uniform distribution on the spin magnitude parallel to the orbital plane, with a peak at 0.

The second *broad* prior allows for a wider range of masses and spins. Here we assume that component masses are uniformly distributed between $0.7 M_\odot$ and $3.0 M_\odot$ with the constraint that $m_2 \leqslant m_1$. Priors on the chirp mass $\mathcal{M}$ and mass ratio $q$ are determined from the concomitant Jacobian transformation. For computational efficiency, we impose a further cut on the chirp mass of the binary, such that we only consider points with chirp mass between 1.190 and 1.210 solar masses. However, this is wide enough to cover the full range of values supported by the data [3, 6, 38, 39]. In the broad prior, the spin magnitudes are taken to be uniformly distributed between 0 and 0.7. This is narrower than the spin prior used in [6] which extended to 0.89. The broad prior is intended to be a conservative spin prior compatible with the fastest-spinning known neutron star, which has a dimensionless spin $\leqslant 0.4$ [69]. The choice of the limit 0.7 should not influence the results, as we obtain one-sided 95% upper limits for the dimensionless spin posterior which are much lower (below 0.35 for all models).

Non-rotating neutron stars cannot exceed a maximum mass which depends on the equation of state. If the sampler chooses a point in the parameter space where a given equation of state cannot support a non-rotating neutron star, we assume that the body is a black hole. Although rotation can increase the maximum mass, we do not allow for a scenario in which any neutron star is so massive that it requires rotation to prevent collapse. This choice is made to ensure that the prior volume is equal for all the equations of state. Our exact model assumption for a given equation of state is thus that objects below the maximum mass of non-rotating neutron stars are in fact neutron stars described by the given equation of state, otherwise light black holes. For the narrow prior, the distinction is irrelevant because we find no posterior support above the maximum mass for any equation of state. For the broad prior on the other hand, the posterior has significant support for mixed binaries for some equation of state models, as is further discussed in section 3.

There is also an upper bound on the dimensionless spin possible for uniformly rotating neutron stars, which depends on the equation of state. The maximum spin covered by the broad prior is below the maximum possible spin for some models considered here (compare section 2.4). Thus, the broad prior will contain a fraction of neutron stars rotating impossibly fast for some equations of state. In practice, this issue is not relevant as we find zero posterior support for such rotation rates.

Finally, we also present results where we compare the hypothesis that both bodies are low-mass black holes against the hypothesis that one, or both, bodies are neutron stars, but with a non-specific equation of state model. In this case we are sampling over the tidal deformability of the neutron stars (and deriving all other equation of state specific quantities from this value as explained below). For this analysis we use the same prior choices as above, specifically considering the broad mass prior, except for the following choices on spins and tidal deformability distributions. Here, a black hole is defined as having a spin magnitude uniformly distributed in $[0, 0.89]$ and zero tidal deformability, while a neutron star has a spin magnitude uniformly distributed in $[0, 0.05]$ and a tidal deformability uniformly distributed in $[0, 5000]$. The upper spin limit for black holes differs from the broad prior considered above. The specific models we study in this case are:

(i) **Binary Black Hole**: $\chi_1 \in [0, 0.89]$, $\chi_2 \in [0, 0.89]$, $\Lambda_1 = \Lambda_2 = 0$.
(ii) **Black Hole—Neutron Star**: $\chi_1 \in [0, 0.89]$, $\chi_2 \in [0, 0.05]$, $\Lambda_1 = 0$, $\Lambda_2 \in [0, 5000]$, $m_{\mathrm{BH}} > m_{\mathrm{NS}}$.
(iii) **Neutron Star—Black Hole**: $\chi_1 \in [0, 0.05]$, $\chi_2 \in [0, 0.89]$, $\Lambda_1 \in [0, 5000]$, $\Lambda_2 = 0$, $m_{\mathrm{NS}} > m_{\mathrm{BH}}$.
(iv) **Binary Neutron Star, no assumptions on equation of state**: $\chi_1 \in [0, 0.05]$, $\chi_2 \in [0, 0.05]$, $\Lambda_1 \in [0, 5000]$, $\Lambda_2 \in [0, 5000]$.
(v) **Binary Neutron Star, neutron stars obey same equation of state**: $\chi_1 \in [0, 0.05]$, $\chi_2 \in [0, 0.05]$, $\Lambda_s \equiv (\Lambda_2 + \Lambda_1)/2 \in [0, 5000]$, $\Lambda_a \equiv (\Lambda_2 - \Lambda_1)/2 = \Lambda_a(\Lambda_s, m_2/m_1)$.

where here $\chi_i$ and $\Lambda_i$ denote the dimensionless spin and tidal deformability of the body $i$ respectively. The analysis of model (iv) allows the two component tidal deformabilities to vary independent of each other. In contrast, the analysis of model (v) uses an equation of state-independent relation between the two tidal deformabilities given the mass ratio of the system [48, 49].

### 2.3. Waveform models for the gravitational-wave signal

Gravitational-wave signals from binary neutron stars are distinct from signals from binary black holes. The deformation of the neutron star due to tidal effects and a spin-induced quadrupole, as well as the post-merger behavior of the remnant, will all leave an imprint in the emitted gravitational waves. Unfortunately, no waveform model yet exists that can consistently model all three of these effects, as modeling the post-merger behavior requires expensive numerical simulations. Such simulations have been performed by a number of groups in recent years, enabling an exploration of the expected features of post-merger emission [70–72]. Though this work has not yet reached the stage where one can predict waveforms with arbitrary masses, spins and equation of state, unmodeled analyses can target the expected post-merger signal [73–75]. Since the post-merger component of the signal occurs at frequencies $\sim$2000 Hz, where Advanced LIGO and Advanced Virgo sensitivity is decreasing, observing this regime was not possible for GW170817 [3, 6, 22, 76]. For the present analysis, it is therefore safe to ignore the post-merger signal.

We use waveform models that predict gravitational-wave signals for a large range of parameters and include both the effects of the tidal deformation and of the spin-induced quadrupole. In contrast to previous analyses [3, 6], where the individual tidal deformabilities $\Lambda$ of the neutron stars were treated as independent intrinsic parameters, we assume a specific equation of state model and the tidal deformability is computed as a function of the masses. To construct the functions $\Lambda(m)$, the Tolman–Oppenheimer–Volkoff differential equations [77, 78] for the stellar structure are solved numerically together with those for the tidal deformability [12, 79].

The quadrupole-monopole effects are caused by the quadrupolar deformations induced by the spin [11], and are quadratic in the spins. These effects are parametrized by a quadrupole-monopole parameter $\kappa$. Similarly to the tidal deformability, this parameter is treated as a pure function of the masses $\kappa(m)$, determined in practice from the value of $\Lambda$ via fits of approximate universal relations for neutron stars [48].

The waveform models used in this study are essentially the same as in the previous analyses from the LIGO–Virgo collaboration [3, 6, 39], to which we refer the reader for more details. The TaylorF2 model uses purely post-Newtonian information to generate closed-form waveforms in the Fourier domain, only allowing for spins aligned with the orbital angular momentum. It includes point-particle and spin contributions (see, e.g. [80]) and tidal contributions. In contrast to [6], we consider several variants of TaylorF2, incorporating tidal terms up to the 6, 7 or 7.5 post-Newtonian order [13, 14, 81][196].

A closed-form tidal approximant, denoted NRTidal, calibrated to binary neutron star numerical relativity simulations was presented in [82, 83]. In this paper, EOB+NRT and Phenom+NRT will denote the waveform models constructed from adding this numerical relativity tuned approximant for tidal effects onto the underlying point-particle waveform models, SEOBNRv4_ROM [84] and IMRPhenomPv2 [85–88] respectively. While the EOB+NRT model is restricted to aligned spins, Phenom+NRT incorporates an effective representation of precession effects [85, 86]. All three models TaylorF2, Phenom+NRT and EOB+NRT

[196] The tidal terms at the 7PN and 7.5PN orders are not strictly complete, with a missing term at 7PN (2PN relative order in the tidal terms) but this missing piece is argued in [14] to be unimportant quantitatively.

include post-Newtonian quadrupole-monopole terms, up to the third PN order in the phasing [11, 89].

In addition, we use two time-domain effective one body models that incorporate tidal effects, SEOBNRv4T [90, 91] and TEOBResumS [92]. SEOBNRv4T includes dynamical tides and the effects of the spin-induced quadrupole moment. TEOBResumS incorporates a gravitational-self-force re-summed tidal potential and the spin-induced quadrupole moment. Both models have been shown to be compatible with state-of-the-art numerical simulations of binary neutron stars up to merger [92, 93]. Also compare [94, 95] regarding recent extensions.

Lacking a prescription for the post-merger signal, all waveforms have an equation of state dependent termination frequency that is close to the merger frequency. For TaylorF2, the termination frequency is determined according to equation (13) of [45][197]. For EOB+NRT and Phenom+NRT, the termination frequency is computed following equation (11) of [83]. Thus the NRT waveforms with zero tidal deformabilities are not strictly speaking equivalent to the binary black hole waveforms: in the NRT prescription the merger is cut instead of terminating with a ringdown signal as would be the case in the underlying point-particle model. However, this difference only concerns high frequencies and should not affect our evidence computations. For most of the configurations considered in this analysis, a remnant black hole would have a ringdown frequency above $\sim$4 kHz, justifying the above statement.

It is important to note that all our waveform models are imperfect and come with systematic errors. Without a quantitative model to represent this as a modeling uncertainty, we cannot marginalize over the waveform error in our analysis. Comparisons between different waveform models [83, 96, 97] indicate that TaylorF2 can produce small differences from Phenom+NRT and EOB+NRT in recovering the tidal parameters, while yielding comparable results for the masses and spins. This was indeed observed in the analysis of GW170817 [3, 6, 38, 39], where the TaylorF2 model yielded larger upper limits for the tidal parameters. In order to obtain a rough estimate how differences between waveform models affect the Bayes factor calculations, we compare results obtained with different waveform families.

### *2.4. Equations of state*

For this work, we have collected 24 different prescriptions describing neutron star equations of state from the literature, computed by various nuclear physics approximations. The full list together with the relevant citations is given in appendix A.

All equations of state describe zero-temperature matter in $\beta$-equilibrium. We require equations of state to respect causality, i.e. the maximum sound speed inside stable neutron stars may not exceed the speed of light at any mass up to the maximum allowed. For comparison with previous studies [6, 39], however, we also include some results for the WFF1 [98] equation of state model, which violates causality before reaching the central density of the maximum-mass neutron star. One equation of state model (HQC18 [9]) incorporates a transition between a hadronic phase in the crust and a quark matter phase in the core. The other equation of state models describe purely hadronic matter without significant phase transitions. We deliberately do not use any constraints from previous studies of the detected event [3, 6, 38, 39] to further narrow the equation of state model selection.

The equation of state properties most important for our analysis of the gravitational-wave signal are the maximum gravitational mass of non-rotating neutron stars and their tidal deformability as a function of gravitational mass. In order to derive implications for the merger remnant, we also require the maximum mass of uniformly rotating neutron stars, as well as the maximum rotation rate and moment of inertia as a function of mass. The maximum dimensionless spin for a given mass is also relevant with regard to the spin prior. We compute all those properties for all equations of state used in this work. Key quantities are tabulated in appendix A.

[197] There is a typo in equation (14) of [45], the denominator should be the sum of the two radii *cubed*, without which the equation is not dimensionally correct.

The properties of non-rotating neutron stars which follow from the equations of state used here are shown in figure 1. For comparison, we also show the credible intervals for the component gravitational masses derived in [6] for different priors on the spin. The maximum gravitational mass covers a range $1.92\,M_\odot$ (BHF_BBB2) to $2.75M_\odot$ (MS1_PP, MS1B_PP). The discovery of neutron star PSR J0348+0432 [26] with mass $2.01 \pm 0.04\,M_\odot$, and the very recent observation of PSR J0740+6620 [99] with mass of $2.14^{+0.10}_{-0.09}\,M_\odot$, provide a lower limit for the maximum mass of neutron stars. However, we also included some models with maximum mass slightly below those constraints in our comparison. Although causality would allow maximum masses even larger than our most massive models (see [100, 101]), the chosen range covers most nuclear physics estimates. The smallest tidal deformability within the mass credible interval [6] for the low spin prior is found for the WFF1 equation of state model, with $\Lambda = 61$, and the largest one for the MS1_PP equation of state model, with $\Lambda = 3728$. In the same mass range, we find that the dimensionless spin at the mass-shedding limit is between 0.66 (KDE0V1) and 0.76 (SKI5).

### 2.5. Gravitational-wave data

For our analysis, we follow [7] and use data from the two LIGO observatories with the full frequency-dependent calibration uncertainties described in [102, 103] and following the subtraction of independently measured noise sources as discussed in [104]. As discussed in [3], a non-Gaussian transient occurs in the data recorded by the LIGO Livingston observatory. This non-Gaussian transient is removed from the data as in [3] using the methods discussed in [73]. We have checked that the instrumental transient subtraction does not bias the results of our inference substantially [105]. We also follow [7] in using data from the Virgo observatory using the calibration and noise-subtraction model discussed in [106]. For both the LIGO and Virgo observatories we analyze data between 23 Hz and 2048 Hz.

### 2.6. Evaluating the Bayes factors

In this work we utilize two different strategies to compute the Bayes factors. We use pre-existing tools in the LALInference [42] and RIFT [43] parameter inference codes to compute the evidence for each equation of state model (equation (2)). For LALInference, this involves performing a distinct detailed Bayesian analysis of the data for each proposed model, using nested sampling [107] and thermodynamic integration [108] to compute the evidence [42]. Both methods have been extensively tested [42] and produce consistent results [109].

For RIFT, this calculation is organized as a two-stage process. The first stage estimates a single marginal likelihood $P(s|x, \Lambda_1, \Lambda_2, I) = \int \mathrm{d}\theta P(s|\theta, x, \Lambda_1, \Lambda_2; M_i, I)P(\theta)$ as a function of the extrinsic parameters $\theta$ and the intrinsic binary parameters $x = (m_1, m_2, \chi_1, \chi_2)$ and $(\Lambda_1, \Lambda_2)$ without assuming an equation of state. The next stage computes the evidence via $P(s|M_i, I) = \int \mathrm{d}xP(x)P(s|x, \Lambda(m_1|M_i), \Lambda(m_2|M_i); M_i, I)$, where $\Lambda(m|M_i)$ denotes the deformability as function of mass assuming the equation of state model $M_i$. Using a precomputed marginal likelihood, this second stage can be performed very rapidly for any equation of state, in seconds to minutes depending on the accuracy required. Unless otherwise noted, all RIFT calculations are performed using the marginalized likelihoods $P(s|x, \Lambda_1, \Lambda_2)$ employed in [7].

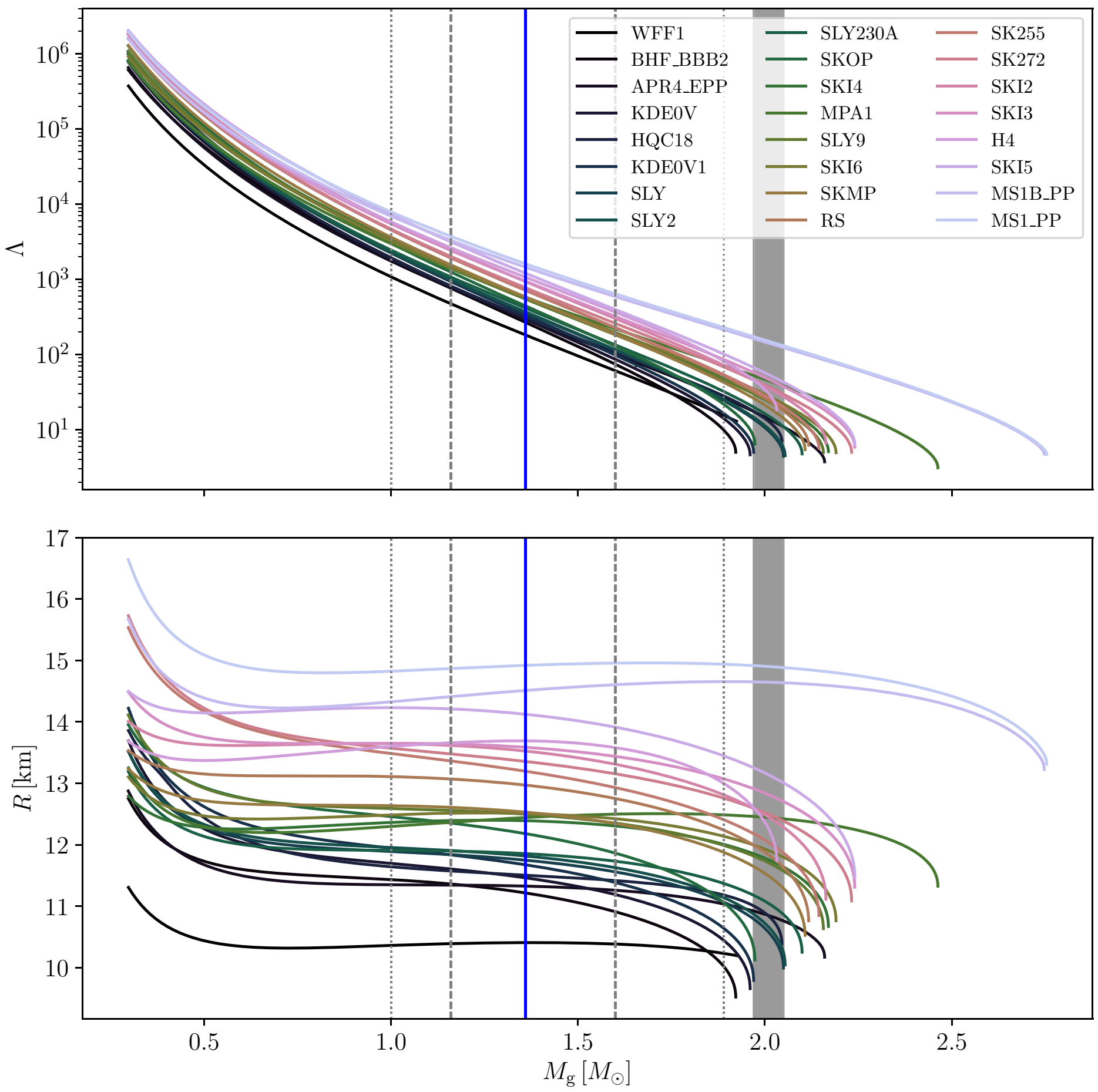

**Figure 1.** Properties of non-rotating neutron stars for all equations of state considered in this work. Top: dimensionless tidal deformability versus gravitational mass. Bottom: proper circumferential radius. All sequences terminate at the maximum mass model (except WFF1, which is cut where the maximum sound speed reaches the speed of light). The legend entries are sorted by the radius of a $1.36\,M_\odot$ neutron star. For comparison, we show the 90% symmetric credible interval given in [6] for the component masses, with dashed vertical lines for the low-spin prior, dotted lines for the high spin prior, and the solid line for the equal-mass limit $1.36\,M_\odot$. The shaded vertical area marks the measured mass for pulsar PSR J0348+0432 [26].

## 3. Results

For each equation of state described in section 2.4, the Bayes factors are computed using either LALINFERENCE or RIFT, using both the narrow and broad priors discussed in section 2.6. The Bayes factors for all waveform approximants and the statistical uncertainties are given in appendix B. The Bayes factors calculated adopting the narrow prior are shown in tables B1 and B3 for LALInference and RIFT, respectively. Similarly, for the broad prior, the LALInference results are given in table B2 and the RIFT results in table B4. The Bayes factors for the LALInference runs are quoted with respect to the low-mass binary black hole hypothesis. The RIFT results re-use the analysis presented in [7] and are not normalized with respect to the BBH case due to the way in which the fits are performed. Instead, the RIFT Bayes factors are quoted relative to the arbitrary hypothesis that GW170817 was a binary neutron star with the SLY9 equation of state, which has modest support and shows relatively little dispersion between waveform approximants in the LALInference results.

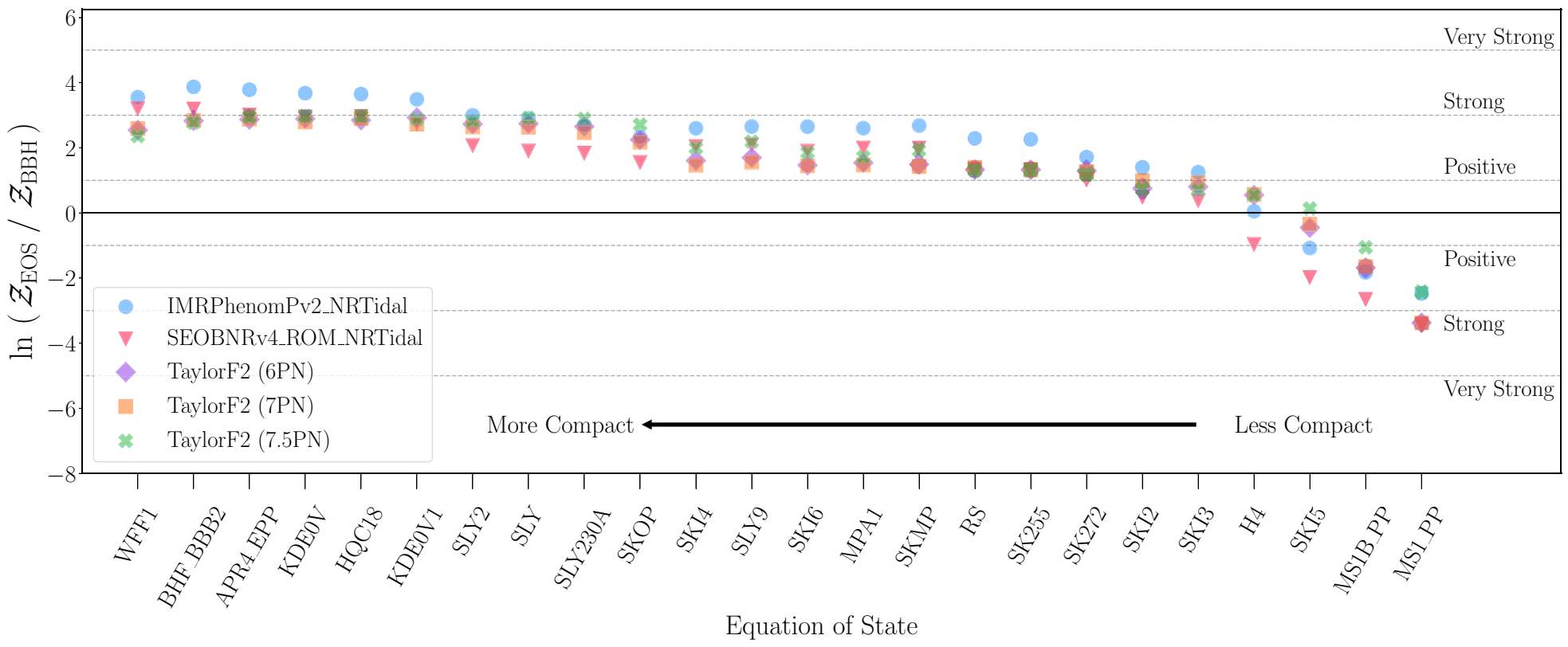

**Figure 2.** The Bayes factors for the narrow prior results using different waveform approximants, as given in table B1. The equations of state have been sorted by the compactness $\mathcal{C}$ of the neutron star at a fiducial mass of $1.36 M_{\odot}$. The results show a clear preference for more compact equations of state. We adopt the guidelines of [111] for the interpretation of the Bayes factors. Here $\mathcal{Z} = P(s|M_i, I)$ is the evidence of a given equation of state or the fiducial binary black hole model. Statistical errors on the Bayes factors from the nested sampling algorithm are negligible.

For the narrow prior, the gravitational-wave data *does not favor and nor can it rule out* any single equation of state with respect to the low-mass binary black hole hypothesis. This can be seen in figure 2, although the data clearly favors softer equations of state, which predict more compact neutron stars with lower tidal deformabilities. The broad prior case, shown in figure 3, exhibits the same trend. The relation is visualized more directly in figure 4, showing fiducial neutron star properties as a function of the Bayes factor with respect to the KDE0V model, which is the most favored for the narrow prior with TaylorF2 (7.5pN) approximant. The models with largest fiducial radius and tidal deformability are *very strongly disfavored* (MS1_PP) or at least *strongly disfavored* (MS1B_PP) compared to the numerous models with small fiducial radius and deformability, such as KDE0V.

Figure 4 also shows upper limits for radius and tidal deformability given by parameter estimation studies [6, 38, 39]. For all our equations of state, and mass ratios in the range $q > 0.7$, the effective tidal deformability and the tidal deformability at the fiducial mass agree better than 6%. To this accuracy, we can compare the upper limits to the values plotted for the fiducial mass models. We find that some of our models with tidal deformabilities above the limits are not strongly disfavored with respect to the most favored equations of state.

Although we can only compute Bayes factors between the models at hand, the Bayes factors exhibit a trend of decreasing with an increasing fiducial radius and tidal deformability. If this trend holds in general, then for a fiducial mass of $1.36 M_{\odot}$ models with tidal deformability $\Lambda(M = 1.36) \gtrsim 1600$ or radius $R(M = 1.36) \gtrsim 14.9\,\mathrm{km}$ are very strongly disfavored.

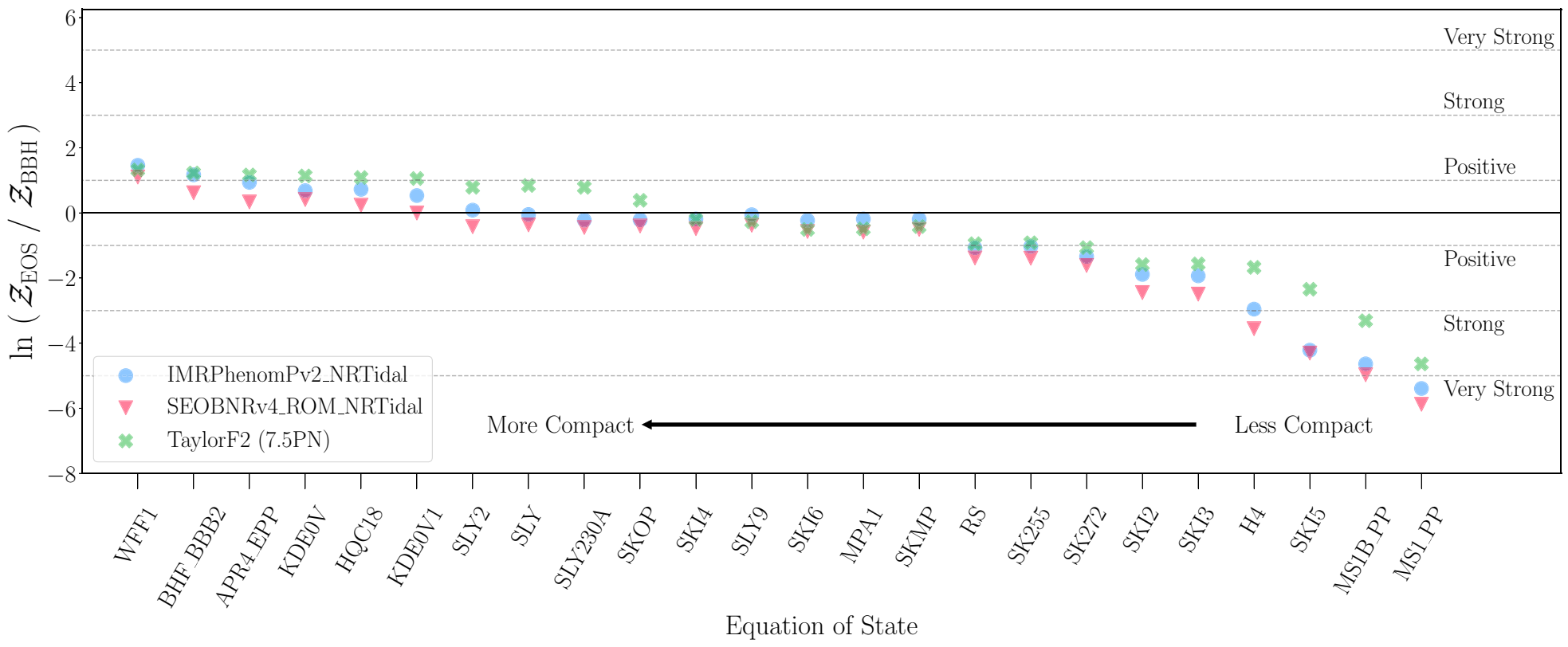

**Figure 3.** As per figure 2 but for the broad prior using different waveform approximants, given in table B2.

For the broad prior, the gravitational-wave data alone does not prefer any single equation of state, as was the case for the narrow prior. However, MS1_PP and MS1B_PP are now *strongly disfavored* against the low-mass binary black hole hypothesis, as shown in figure 3. Normalizing the Bayes factors against KDE0V, we find that MS1_PP is *very strongly* and MS1B_PP *strongly* disfavored with respect to the softer KDE0V equation of state, as shown in figure 4.

In contrast to the narrow prior, we find that for some equations of state there is significant posterior support for a mixed binary where the heavier object is a light black hole. The reason for this is that the mass range is less constrained and exceeds the maximum possible mass of non-rotating neutron stars for those equations of state. Models BHF_BBB2, KDE0V1, HQC18, and KDE0V have 13, 7, 5, and 7 percent posterior probability, respectively, for a mixed binary. The remaining models have support below 5% for mixed binary, in particular MS1_PP and MS1B_PP have zero support for mixed binaries. For the broad prior and the WFF1 equation of state, there is a $\approx$14% posterior probability for a central neutron star density in an unphysical (causality violating) region of the parameter space.

For both sets of priors, we find that the bulk of equations of state considered in this paper produce evidences that are consistent to within an order of magnitude. This is in line with the expectations for second-generation gravitational-wave detectors, where $\sim\mathcal{O}(20)$ detections will allow hypothesis ranking to distinguish between hard, moderate and soft equations of state [44]. Even if the true equation of state were to be one of those considered in this paper, we should not necessarily expect this to be the most preferred equation of state in the hypothesis ranking [45]. For example, this could arise due to the effects of noise or other systematics. More realistically, it is also possible that the true equation of state is not contained in the finite set of models considered here, though we expect the highest ranked hypotheses to share similar macroscopic features to the true neutron star equation of state.

It is important to assess the systematic bias inherent in our waveform models. Such systematic biases could arise from a different treatment of the underlying point particle limit, which can start to become significant at higher frequencies where tidal effects also become more prominent. Additionally, the different waveform models employ different prescriptions for the tidal contribution to the phasing, which can lead to discrepancies in the recovery of tidal parameters. As demonstrated in previous studies [39, 83, 96, 97, 110], the tidal effect calculated in the TaylorF2 waveform models is typically smaller than that predicted using the NRTidal prescription. As such, the tidal parameters inferred from the data using the TaylorF2 waveform model will typically be larger in order to compensate for the reduced tidal effect. Consequentially, we find that the Bayes factors calculated using the TaylorF2 model tend to show greater support for stiffer equations of state, which yield larger tidal deformabilities, in comparison to the effective-one-body or phenomenological waveform models. However, as can be seen in figures 2–4, the Bayes factors between the different waveform approximants still show reasonably good agreement.

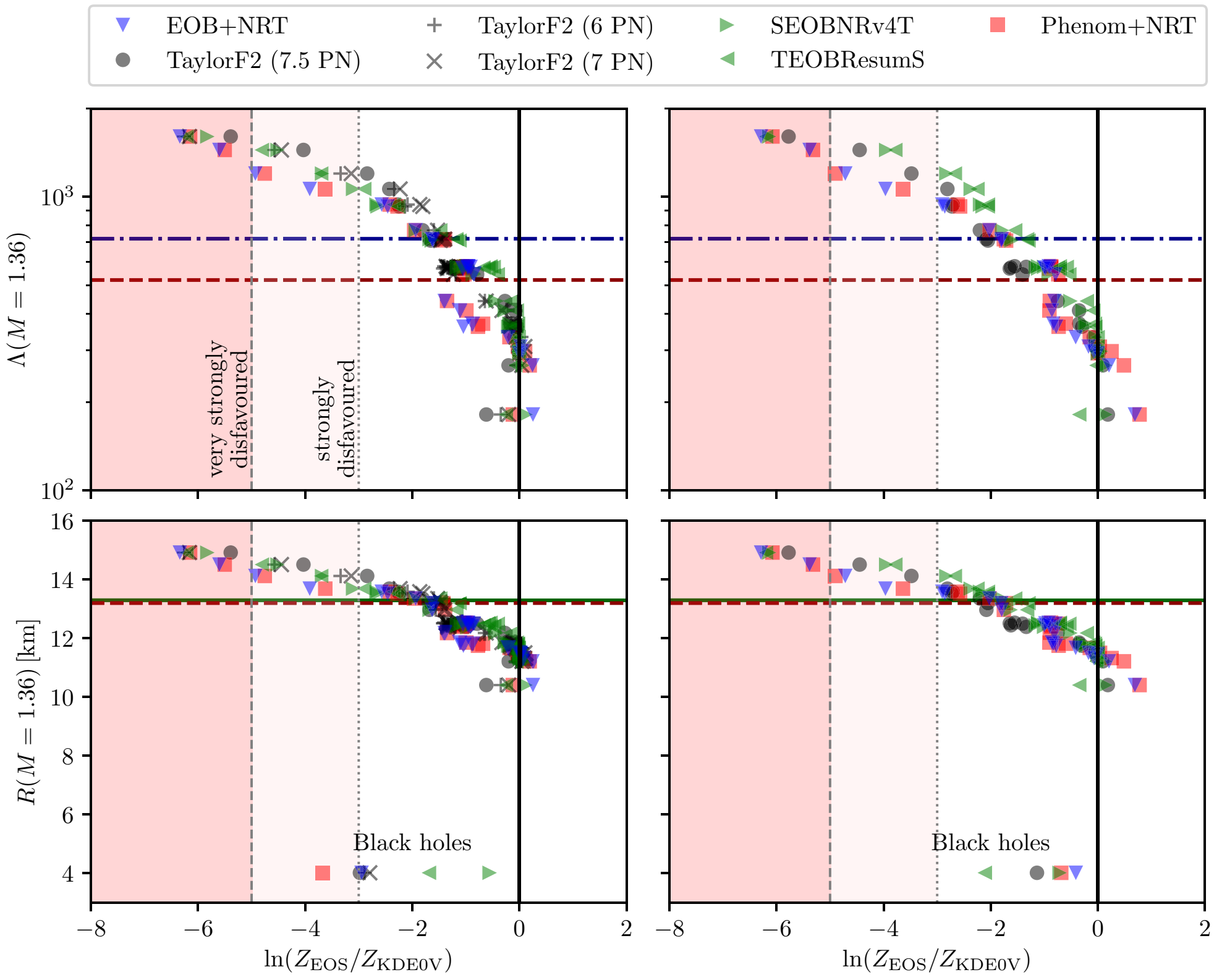

**Figure 4.** The fiducial tidal deformability $\Lambda$ at $1.36M_\odot$ (top) and the fiducial neutron star radius $R$ at $1.36M_\odot$ (bottom) plotted as a function of the Bayes factors for the narrow prior (left) and broad prior (right). For the binary black hole case, the plotted radius is the Schwarzschild radius. The Bayes factors have been normalized with respect to KDE0V, which is the most favored equation of state for the narrow prior with TaylorF2 (7.5pN) waveform model. Compared to the KDE0V model, MS1_PP is very strongly disfavored ($\ln \mathrm{BF} \leqslant -5$) and MS1B_PP is strongly disfavored ($\ln \mathrm{BF} \leqslant -3$) for both priors and all waveform approximants. For comparison, we plot limits from the literature. The solid green horizontal lines mark the upper limit for the coalescing neutron star radii from [39]. Red dashed lines mark the upper limit given for the radius given in [38] (assuming a common radius for the two stars) and the one-sided upper limit for the effective tidal deformability from [38]. The blue dash-dotted lines mark the upper limit (highest posterior density interval) for the effective tidal deformability from [6].

Finally, we also ask if can we rule out the binary black hole hypothesis by computing Bayes factors for the binary neutron star hypothesis, or the hypothesis that there was only one neutron star in the binary, without assuming a specific equation of state. We consider both the hypotheses that the single neutron star is heavier and lighter than its potential black hole companion. The specific priors that we use in these cases were discussed earlier in section 2.2. Table 1 presents the Bayes factors compared to the 'binary neutron star with independent equation of state' hypothesis. We present results with different waveform approximants and methodologies to compute the evidence of each model.

**Table 1.** Bayes factors between various hypotheses on the nature of the GW170817 components obtained using different waveform approximants and two methods for computing the evidence. All Bayes factors are given in relation to the 'binary neutron star with independent equations of state' hypothesis. We do not present results for Phenom+NRT/Nest when a BH is part of the binary, as the combination of precession and high spins makes this analysis computationally very expensive. We find no clear preference for any hypothesis. The table abbreviations stand for nested sampling (Nest), Markov-chain Monte-Carlo/thermodynamic integration (MCMC), binary black hole (BBH), mixed binary where the neutron star is heavier (NSBH), mixed binary where the black hole is heavier (BHNS), binary neutron star (BNS), equation of state (EoS).

| Waveform/Method | BBH | NSBH | BHNS | BNS (common EoS) |
|---|---|---|---|---|
| Phenom+NRT/Nest | — | — | — | 5.0 |
| Phenom+NRT/MCMC | 1.1 | 0.7 | 1.7 | 9.6 |
| EOB+NRT/MCMC | 1.7 | 1.6 | 4.0 | 5.5 |
| TaylorF2 (7pN)/Nest | 1.9 | 1.4 | 3.7 | 4.9 |
| TaylorF2 (7pN)/MCMC | 1.0 | 1.2 | 3.4 | 4.4 |
| TaylorF2 (7.5pN)/Nest | 1.6 | 1.3 | 3.2 | 4.4 |
| TaylorF2 (7.5pN)/MCMC | 0.9 | 1.0 | 5.5 | 5.0 |

From the results in table 1 we can draw a number of conclusions. First, we find that the computed Bayes factors agree closely between different waveform approximants, indicating that uncertainties in the waveform modeling are not compromising our results. Second, we find that both nested sampling and thermodynamic integration return consistent results for the Bayes factors, giving us confidence on the validity of the calculation. Third, and most interesting from a physical perspective, we find that the GW170817 data alone does not strongly favor one hypothesis over the other. All of the Bayes factors we compute are in the range of 1 to 5, suggesting that the gravitational-wave data alone do not show a strong preference for any of the hypotheses considered here.

It may seem strange that the binary neutron star model, assuming no common equation of state, is the least favored hypothesis. Bayes factors depend on the Occam penalty, i.e. how well a model fits the given data (goodness-of-fit) and how many constrained parameters a model includes (dimensionality). We expect all models to be preferred over the binary neutron star model, assuming a common equation of state, if they fit the data equally well. The binary black hole model has two fewer parameters and the neutron star-black hole hypotheses have one fewer parameter. A brief back-of-the-envelope calculation can be used to support the Bayes factors seen in table 1. The Occam penalty is given by $\delta\theta/\Delta\theta$, the posterior width over the prior width, and penalizes a model for each additional constrained parameter it employs, since $\delta\theta < \Delta\theta$. From section 2.2, we see that the width of the prior on the tidal parameters is $\Delta\Lambda \sim 5000$ whereas the width on the posterior is on the order of $\delta\Lambda \sim 1000$ [6]. From this, we expect the binary neutron star hypothesis with a common equation of state to be preferred over an independent equation of state hypothesis by a factor ~5, as it uses one fewer parameter. This is in agreement with the results found in table 1 and as demonstrated in [49] for simulated binary neutron star signals.

## 4. Implications

After comparing the relative probabilities of the different equation of state models, we now discuss the implications that follow from assuming any of the equations of state are correct. We also discuss constraints for the maximum mass of neutron stars that follow from assumptions on the fate of the remnant, but without assuming any equation of state.

### *4.1. Baryonic mass and central density*

The total baryonic mass (defined here as baryon number times a fiducial mass constant $1.66 \times 10^{-27}$ kg) is the most important quantity for the evolution during the post-merger phase, including the lifetime of the remnant and the mass of the accretion disk, which are relevant for the short gamma-ray burst and the kilonova. The maximum density inside the initial neutron stars is important to interpret our results properly. The relative probabilities we compute for different equation of state models only signify that *in the density range occurring in the coalescing neutron stars*, matter is more likely described by one equation of state model than another.

To compute the baryonic mass, we start with the posterior samples obtained assuming a given equation of state model. From the gravitational masses and spins in a given sample, we obtain the baryonic masses of the neutron stars. Those are computed using interpolation tables we build for sequences of non-rotating and maximally rotating neutron stars for each equation of state model. To account for the spin, we expand the gravitational mass at given baryonic mass to second order in the squared dimensionless spin parameter. The zeroth- and first-order coefficients are determined from the binding energy and moment of inertia of the non-rotating model. The second-order coefficient is then determined from the angular momentum and gravitational mass of the maximally rotating model. The spin corrections are relatively small compared to the statistical error both for narrow and broad priors. To obtain limits for the energy densities that can be reached before merger, we use the central densities of non-rotating neutron stars with baryonic masses computed as described above.

The results for the narrow prior are given in table 2. We find that the pre-merger gravitational-wave signal does not carry information about matter at energy densities above $0.5\text{–}1.3 \times 10^{15}$ g cm$^{-3}$, assuming the true equation of state is similar to one of the models considered here. This agrees well with the results by [39] (figure 2 therein) obtained using parametrized equations of state, and also with results in [40] based on Gaussian process models for the equation of state.

### *4.2. Fate of merger remnant*

Next, we assess the fate of the merger remnant, which could either form a black hole directly at merger (prompt collapse), a short-lived hypermassive neutron star, a long-lived supramassive neutron star, or even an unconditionally stable neutron star. This is particularly relevant with regard to the observed short gamma-ray burst and kilonova counterparts. One possible engine for short gamma-ray bursts is given by a black hole submerged in a massive disk [52]. Numerical simulations suggest that prompt black hole formation at merger might be in tension with the production of a short gamma-ray bursts [29, 112], favoring delayed collapse. The other model for the engine consists of a rapidly rotating and strongly magnetized neutron star (magnetar scenario [53, 113, 114]) embedded in a disk. In this model, the remnant needs to survive at least long enough to initiate a jet (however, there can be a delay until the jet produces the observed short gamma-ray burst). The fate of the remnant is also tightly related to the ejection of matter and the spectral evolution of the resulting kilonova [5, 115–118]. For example, the amount, composition and velocity of ejected matter required for the observed kilonova might be in tension with expectations for the prompt collapse case (for a recent review on multi-messenger constraints, see [119]). However, the theoretical modeling of the kilonova is a very complex and rapidly evolving field of research (e.g. [120–123] and the references therein).

**Table 2.** Implications derived from the equation of state model specific parameter estimation runs with narrow prior and TaylorF2 (7.5 PN) waveform model. $M_B^0$ is the total initial baryonic mass. We provide the median value and the 90% two-sided credible bounds. $E_c$ is the 90% one-sided upper limit for the central energy density of the heavier neutron star. The normalized mass $\tilde{m}_0$ is simply a linear transform of $M_B^0$ defined by equation (4), such that the supramassive mass range for the given equation of state model corresponds to $\tilde{m} \in (0, 1)$. The potential impact of mass ejection is given in terms of $\Delta_{0.1} = \tilde{m}_0 - \tilde{m}_R$, computed from equation (5) for a fiducial ejecta mass 0.1 $M_\odot$ (the given value is the 90% one-sided upper limit). The 'prompt collapse' column denotes whether a black hole can be formed directly during the merger, based on available estimates of the corresponding mass threshold, and assuming prompt collapse only occurs for hypermassive systems. Hypermassive cases lacking data on the prompt collapse threshold are denoted with a dash, while 'possibly' refers to the case where the available bounds on the threshold are compatible with both outcomes. $I_{max}$ and $F_{max}$ are upper limits (see main text) for moment of inertia and rotation rate for a uniformly rotating remnant. $I_{max}$ is given in geometric units of $4.335\,871 \times 10^{43}$ g cm$^2$.

| Equation of state | $M_B^0$ ($M_\odot$) | $E_c/c^2$ ($10^{15}$ g cm$^{-3}$) | $\tilde{m}_0$ | $\Delta_{0.1}$ | Prompt collapse | $I_{max}$ ($M_\odot^3$) | $F_{max}$ (kHz) |
|---|---|---|---|---|---|---|---|
| APR4_EPP | $3.021^{+0.019}_{-0.009}$ | 1.05 | $0.883^{+0.041}_{-0.019}$ | 0.22 | No | 84 | 1.79 |
| MS1_PP | $2.940^{+0.014}_{-0.008}$ | 0.47 | $-0.608^{+0.023}_{-0.013}$ | 0.17 | No | 153 | 1.00 |
| MS1B_PP | $2.949^{+0.014}_{-0.008}$ | 0.49 | $-0.581^{+0.023}_{-0.013}$ | 0.16 | No | 152 | 1.03 |
| KDE0V | $3.013^{+0.019}_{-0.009}$ | 1.17 | $1.859^{+0.050}_{-0.025}$ | 0.27 | — | 66 | 1.86 |
| SKI6 | $2.994^{+0.016}_{-0.009}$ | 0.82 | $0.848^{+0.036}_{-0.018}$ | 0.22 | No | 97 | 1.53 |
| H4 | $2.978^{+0.015}_{-0.009}$ | 0.65 | $1.385^{+0.035}_{-0.017}$ | 0.22 | No | 104 | 1.36 |
| MPA1 | $3.004^{+0.016}_{-0.009}$ | 0.74 | $-0.016^{+0.029}_{-0.016}$ | 0.18 | No | 114 | 1.34 |
| SK255 | $2.968^{+0.015}_{-0.008}$ | 0.80 | $1.120^{+0.038}_{-0.020}$ | 0.24 | — | 91 | 1.61 |
| SKI2 | $2.968^{+0.016}_{-0.008}$ | 0.72 | $1.025^{+0.036}_{-0.019}$ | 0.23 | — | 100 | 1.53 |
| SKI4 | $2.997^{+0.017}_{-0.008}$ | 0.83 | $0.911^{+0.035}_{-0.018}$ | 0.22 | No | 94 | 1.58 |
| SKOP | $2.994^{+0.016}_{-0.009}$ | 1.02 | $1.808^{+0.043}_{-0.022}$ | 0.26 | — | 73 | 1.70 |
| SLY2 | $3.005^{+0.018}_{-0.009}$ | 1.02 | $1.401^{+0.043}_{-0.022}$ | 0.25 | — | 77 | 1.79 |
| SLY9 | $2.988^{+0.016}_{-0.008}$ | 0.86 | $0.999^{+0.037}_{-0.019}$ | 0.23 | — | 91 | 1.68 |
| BHF_BBB2 | $3.021^{+0.020}_{-0.009}$ | 1.26 | $2.032^{+0.053}_{-0.025}$ | 0.27 | — | 63 | 1.87 |
| HQC18 | $3.014^{+0.018}_{-0.010}$ | 1.03 | $1.305^{+0.040}_{-0.020}$ | 0.22 | — | 81 | 1.72 |
| KDE0V1 | $3.005^{+0.018}_{-0.009}$ | 1.13 | $1.837^{+0.048}_{-0.024}$ | 0.27 | — | 68 | 1.80 |
| RS | $2.980^{+0.016}_{-0.008}$ | 0.80 | $1.174^{+0.037}_{-0.019}$ | 0.24 | — | 92 | 1.61 |
| SK272 | $2.965^{+0.016}_{-0.008}$ | 0.75 | $0.780^{+0.036}_{-0.019}$ | 0.23 | No | 100 | 1.43 |
| SKI3 | $2.967^{+0.015}_{-0.008}$ | 0.68 | $0.744^{+0.034}_{-0.018}$ | 0.22 | No | 107 | 1.35 |
| SKI5 | $2.957^{+0.015}_{-0.008}$ | 0.63 | $0.775^{+0.033}_{-0.019}$ | 0.23 | No | 109 | 1.32 |
| SKMP | $2.990^{+0.016}_{-0.008}$ | 0.86 | $1.186^{+0.037}_{-0.019}$ | 0.23 | — | 88 | 1.68 |
| SLY | $3.005^{+0.017}_{-0.009}$ | 1.04 | $1.425^{+0.042}_{-0.021}$ | 0.25 | Possibly | 76 | 1.82 |
| SLY230A | $3.007^{+0.017}_{-0.009}$ | 0.97 | $1.175^{+0.040}_{-0.020}$ | 0.23 | — | 83 | 1.76 |

The most relevant quantity regarding the remnant type is the baryonic mass $M_B^R$ of the system, in comparison to the maximum baryonic mass of non-rotating neutron stars, $\hat{M}_B^{TOV}$, and the maximum baryonic mass of neutron stars uniformly rotating at the mass-shedding limit, $\hat{M}_B^{ROT}$. Since the baryon number is conserved, the remnant mass is given by $M_B^R = M_B^0 - M_B^E$, where $M_B^E$ denotes the baryonic mass expelled from the system dynamically and by winds, and $M_B^0$ the total initial baryonic mass.

For discussing the expected remnant evolution, we introduce a useful dimensionless measure $\tilde{m}$ defined as

$$\tilde{m}(M_B) = k_m \left( \frac{M_B}{\hat{M}_B^{TOV}} - 1 \right), \qquad k_m = \frac{\hat{M}_B^{TOV}}{\hat{M}_B^{ROT} - \hat{M}_B^{TOV}}. \tag{4}$$

The maximum masses $\hat{M}_B^{ROT}$ and $\hat{M}_B^{TOV}$, which we compute for each equation of state model are given in table A1. Based on $\tilde{m}_R \equiv \tilde{m}(M_B^R)$ for a given equation of state model, we classify the remnant as hypermassive if $\tilde{m}_R > 1$, supramassive if $0 \leqslant \tilde{m}_R \leqslant 1$, and stable if $\tilde{m}_R < 0$.

The quantity $\tilde{m}_R$ serves as a rough indication for the remnant lifetime $\tau_R$. A stable remnant will never form a black hole, a hypermassive remnant either forms a black hole directly at merger or collapses within a few hundred ms. Within the narrow supramassive mass range, the lifetime decreases sharply. Computing $\tau_R$ as a function of $\tilde{m}_R$, however, is a difficult problem beyond the scope of this paper. Although we expect $\tilde{m}_R$ to be the most important parameter, note that $\tau_R$ might also depend on other factors such as mass ratio, spins, and equation of state.

In table 2, we provide the value $\tilde{m}_0$ corresponding to the median value of the total initial baryonic mass inferred for the narrow prior. Applying equation (4) to $M_B^R$ and $M_B^0$, we obtain

$$\tilde{m}_R = \tilde{m}_0 - (\tilde{m}_0 + k_m) \frac{M_B^E}{M_B^0}. \tag{5}$$

In [116], the ejecta mass was estimated from the observed kilonova to around $0.078\,M_\odot$. If we assume more conservatively that $M_B^E < 0.1\,M_\odot$, we find that $0 < \tilde{m}_0 - \tilde{m}_R < 0.3$ for the equations of state considered here (see table 2). The above ejecta mass limit should be regarded as a fiducial value since modeling the ejecta mass, either by numerical simulations or inferred from the kilonova, is an ongoing field of research.

For the narrow prior, we find that the MS1_PP, MS1B_PP, and MPA1 equation of state models lead to a stable remnant and are therefore incompatible with the black hole with disk short gamma-ray burst scenario (technically, the MPA1 also allows marginally supramassive remnants, which can for all practical purposes be considered stable). The H4, SLY, KDE0V, KDE0V1, SKOP, SLY2, BHF_BBB2, and HQC18 equation of state models result in a short-lived hypermassive remnant. For APR4_EPP, RS, SLY9, SLY230A, SKI2, SKI4, SK255, and SKMP, the remnant is in the upper supramassive or lower hypermassive mass range, and we cannot predict if the lifetime is longer or shorter than the observed short gamma-ray burst delay. For SKI3, SKI5, SKI6, and SK272 the remnant mass is in the supramassive range with $\tilde{m}_R < 0.9$, and likely long-lived.

We now turn to discuss whether the merger can lead to prompt black hole formation for a given equation of state model, based on gravitational-wave data alone. Prompt collapse occurs above some critical mass, which depends on the equation of state. Known examples of prompt collapse in numerical binary neutron star simulations generally involve hypermassive systems. If a parameter estimation run for a given equation of state model yields posterior support only in the supramassive and stable mass range, we assume that no prompt collapse can occur. For hypermassive systems, we use existing thresholds [18] from numerical simulations, given in terms of total gravitational mass of the binary. We consider a systematic error of $0.1\,M_\odot$ due to the granularity of the simulated systems. The available thresholds are for equal-mass, non-spinning systems. To account for a possible impact of mass ratio and spin, we further increase the error by $0.05\,M_\odot$. This is a conservative estimate assuming that the main influence is by the change in dynamically ejected mass, which has been computed for many models in [124]. Finally, we express the thresholds for equal mass systems in terms of total baryonic mass and use it as an approximate threshold on $M_{\mathrm{B}}^{\mathrm{tot}}$ for the generic case.

For the narrow prior, we can rule out prompt collapse for the APR4_EPP, H4, MPA1, MS1_PP, MS1B_PP, SKI3, SKI4, SKI5, SKI6, and SKI272 equation of state models. The SLY equation of state model is compatible both with prompt and delayed collapse (due to the systematic error of the threshold mass from numerical relativity). For the remaining equations of state, no data on prompt collapse threshold is available.

With improved theoretical modeling, the wealth of electromagnetic observations may yield reliable constraints on the lifetime of the remnant. If prompt collapse could be ruled out, the values of $M_{\mathrm{B}}^{0}$ obtained from gravitational-wave data could be taken as a lower limit for the prompt collapse threshold. Considering a potential impact of the mass ratio on the actual threshold, one has to take into account that the limits on $M_{\mathrm{B}}^{0}$ are the result of a marginalization over mass ratio.

### *4.3. Constraining the maximum neutron star mass*

Within the black hole with disk scenario of short gamma-ray bursts, the remnant cannot be stable. This was already exploited in [4] to derive limits on the maximum mass of non-rotating neutron stars. The delay of 1.7 s [4] between merger and short gamma-ray burst would also constrain the lifetime of the remnant. The relatively short lifetime indicates that the remnant either cannot be supported by uniform rotation alone (hypermassive remnant), or that it is at least close to the critical mass (how close exactly is an important open question).

In the following, we estimate the maximum mass of non-rotating neutron stars under the assumption that a black hole was produced in the binary neutron star merger before the observed short gamma-ray burst. For this, we first rewrite equation (4) as

$$\hat{M}_{\mathrm{B}}^{\mathrm{TOV}} = M_{\mathrm{B}}^{0}\frac{1 - M_{\mathrm{B}}^{\mathrm{E}}/M_{\mathrm{B}}^{0}}{1 + \tilde{m}_{\mathrm{R}}/k_m} \leqslant \frac{M_{\mathrm{B}}^{0}}{1 + \tilde{m}_{\mathrm{R}}/k_m}. \tag{6}$$

From table 2, we find that $M_{\mathrm{B}}^{0}$ varies only slightly with equation of state. Therefore, we take the bound $M_{\mathrm{B}}^{0} < 3.05\,M_\odot$ for the equations of state considered here as a generic bound. In addition, we assume $k_m < 7$, as fulfilled by all equations of state considered here. The remaining open question is below which value $\tilde{m}_{\mathrm{R}}$ the remnant lifetime is above the short gamma-ray burst delay. We already know that the lifetime is infinite unless $\tilde{m}_{\mathrm{R}} > 0$. Future studies might provide tighter bounds, which will tighten the upper bound obtained for $\hat{M}_{\mathrm{B}}^{\mathrm{TOV}}$. For example, if we demand $\tilde{m}_{\mathrm{R}} > 0.4$, then $\hat{M}_{\mathrm{B}}^{\mathrm{TOV}} < 2.9\,M_\odot$.

Another potential method to limit the maximum mass of non-rotating neutron stars is to limit the angular momentum that needs to be removed before collapse can occur, and which becomes zero for $\tilde{m}_{\mathrm{R}} = 1$. In [27], it was argued that the remnant mass must be close to the hypermassive range since the large spindown energy that would otherwise have been deposited in the environment would be incompatible with the observed counterparts (however, there are observational gaps [5], e.g. in x-rays, which could lead to energy output being missed). This assumption corresponds to $\tilde{m}_{\mathrm{R}} \geqslant 1$. From equation (6), we would obtain a corresponding limit $\hat{M}_{\mathrm{B}}^{\mathrm{TOV}} < 2.67\,M_\odot$ (for the narrow prior). If we further assume that the maximum baryonic mass is 15%–23% larger than the maximum gravitational mass, as fulfilled for all equation of state models considered here, we obtain a limit $\hat{M}_G^{\mathrm{TOV}} \leqslant 2.32\, M_\odot$. Based on additional assumptions, excluding prompt collapse and using an approximate expression for the prompt collapse threshold ignoring systematic errors, [27] provides a limit on the maximum neutron star mass which is 6% tighter. Reliable prompt collapse thresholds for a representative selection of equation of state models might hence be useful in this respect.

In [29], a limit $\hat{M}_G^{\mathrm{TOV}} \leqslant 2.28 \pm 0.23\, M_\odot$ was provided, again based on the assumption of having a hypermassive remnant. The differences to our analysis are mainly the use of gravitational masses instead of conserved baryonic masses, and a more constraining assumption regarding the range of ratios between maximum gravitational masses for uniformly rotating and non-rotating neutron stars. Such effects are included in the systematic error given in [29] for their maximum mass limit, which is therefore compatible with our result.

A similar calculation can be found in [54], yielding a limit $\hat{M}_G^{\mathrm{TOV}} \leqslant 2.15\text{–}2.25\, M_\odot$. It is based again on the detected limits for the total initial gravitational mass, and further assumes a total energy loss of $(0.15 \pm 0.03)\, M_\odot$, as well as a fixed difference $0.4\, M_\odot$ between maximum gravitational masses of uniformly rotating and non-rotating neutron stars. These assumptions might explain the tighter limit compared to our result (for example, we find $\hat{M}_G^{\mathrm{ROT}} - \hat{M}_G^{\mathrm{TOV}} = 0.34\, M_\odot$ for the BHF_BBB2 equation of state). A recent analysis [125], which is not relying on assumptions about $\tilde{m}$, indicates that the remnant at the time of collapse might not rotate rapidly, and provides a limit $\hat{M}_G^{\mathrm{TOV}} \leqslant 2.3\, M_\odot$.

By using the detected gravitational-wave signal alone, without assumptions on the remnant fate, [40] inferred a limit $\hat{M}_G^{\mathrm{TOV}} \leqslant 2.46$, compatible with our results. This is based on modeling the equation of state using Gaussian processes, trained on a set of seven equations of state and respecting causality by construction. The Gaussian process model is further informed by the gravitational-wave data and then used to *extrapolate* the equation of state from densities relevant for the inspiral signal up to those required to determine the maximum mass.

### *4.4. Impact of prior and waveform model*

In order to assess the waveform systematics, we repeat the computation of the remnant properties for a representative equation of state model using all waveform approximants. The results are shown in table 3. The differences for $M_B$ and $\tilde{m}_0$ are small compared to the statistical uncertainties, and negligible with regard to our conclusions on remnant classification, prompt collapse, and the maximum neutron star mass. This is to be expected, since those results depend directly only on the initial neutron star masses, but not on the tidal deformabilities, which are most sensitive to the waveform model.

Table 4 shows the results obtained using the broad prior restricted to the binary neutron star case. The restriction is necessary as the heavier object exceeds the maximum mass allowed for a neutron star for parts of the broad prior. Comparing to the narrow prior in table 2, we find that the total initial baryonic mass distributions are broadened towards higher masses. This is a consequence of the wider posterior distribution for the mass ratio; for fixed chirp mass, the total baryonic mass for unequal mass binaries is higher than for equal masses.

The broader distribution leads to differences regarding the fate of the remnant. While we ruled out prompt collapse for the H4 and APR4_EPP equations of state with the narrow prior, the respective total mass posteriors for the broad prior also have support in the mass range where prompt collapse is allowed according to the thresholds from numerical relativity. This result might still change with more accurate numerical-relativity results for the threshold, given its large systematic error.

**Table 3.** Implications derived assuming the MPA1 equation of state model and narrow prior, using different waveform approximants. The given quantities are explained in table 2.

| Waveform | $M_B$ ($M_\odot$) | $E_c/c^2$ ($10^{15}$ g cm$^{-3}$) | $\tilde{m}_0$ | $\Delta_{0.1}$ | $I_{max}$ ($M_\odot^3$) | $F_{max}$ (kHz) |
|---|---|---|---|---|---|---|
| TaylorF2(7.5 PN) | $3.004^{+0.016}_{-0.009}$ | 0.74 | $-0.016^{+0.029}_{-0.016}$ | 0.18 | 114 | 1.34 |
| TaylorF2(7 PN) | $3.004^{+0.016}_{-0.009}$ | 0.74 | $-0.017^{+0.029}_{-0.015}$ | 0.18 | 114 | 1.34 |
| TaylorF2(6 PN) | $3.004^{+0.016}_{-0.009}$ | 0.74 | $-0.016^{+0.028}_{-0.016}$ | 0.18 | 114 | 1.34 |
| EOB+NRT | $3.004^{+0.016}_{-0.008}$ | 0.74 | $-0.015^{+0.027}_{-0.015}$ | 0.18 | 114 | 1.34 |
| Phenom+NRT | $3.005^{+0.016}_{-0.009}$ | 0.74 | $-0.015^{+0.029}_{-0.015}$ | 0.18 | 114 | 1.34 |

The KDE0V, SKI3, SKI4, SKI5, SKI6, and SK272 equations of state, which lead to supramassive remnants assuming the narrow prior, might also lead to hypermassive remants when assuming the broad prior. The MS1_PP, MS1B_PP equations of state still lead to stable remnants, while the MPA1 equation of state model can also result in a long-lived supramassive remnant for the broad prior. The bound on the maximum Tolman–Oppenheimer–Volkoff mass when assuming the black hole with disk short gamma-ray burst model becomes less strict as well. Using only $\tilde{m}_R \geqslant 0$, we obtain $\hat{M}_B^{TOV} < 3.4 M_\odot$. Assuming $\tilde{m}_R \geqslant 1$, we would obtain $\hat{M}_B^{TOV} < 2.98 M_\odot$.

Another consequence of the larger mass ratios is that the mass posterior of the heavier neutron star extends to larger values. Therefore, higher central energy densities are reached. Depending on the equation of state, the inspiral gravitational-wave signal might be sensitive to the equation of state at energy densities up to $2 \times 10^{15}$ g cm$^{-3}$ (compare table 4).

### *4.5. Remnant rotation rate*

In case the remnant is long-lived, the maximum rotation rate and moment of inertia are of interest for searches of long-lived gravitational-wave signals from the remnant (see [20, 22]), and might also be relevant for modeling the magnetar short gamma-ray burst scenario. In the following, we estimate upper limits for rotation rate and moment of inertia.

The total angular momentum at merger is not known, although numerical simulations indicate that it typically exceeds the maximum amount that could be realized in a uniformly rotating system of the same mass (see [21]). However, the disk can account for a significant fraction of the total angular momentum. Most of the disk angular momentum is transported outwards during accretion (together with a fraction of the mass). The angular momentum of the final remnant could depend on the details of the accretion process and might be less than the critical one (the opposite assumption is made in [21]), and it can decrease further on longer timescales, e.g. by magnetic spindown. The rotation rate of models near the maximum mass can increase with decreasing angular momentum, because the compactness also increases. The maximum rotation rate can in principle exceed the rotation rate at the mass-shedding limit. Here, we use the rotation rate at the mass-shedding limit as an estimate for the maximum rotation rate of the remnant.

From the mass posteriors for each equation of state model, we compute the posteriors for the rotation rate and moment of inertia of uniformly rotating neutron stars with baryonic mass equal to the total baryonic mass and maximum angular momentum. Tables 2 and 4 show the numerical results for the narrow and broad priors. For both, we find a maximum rotation rate below 1.9 kHz for all equations of state. We also provide values for remnants in the hypermassive mass range. For those cases, we limit the baryonic mass above to the maximum allowed for uniform rotation. The numbers we obtain can be regarded as estimates for the hypothetical case that the excess mass was somehow lost during merger, or remains in a disk.

**Table 4.** Like table 2, but showing results for the broad prior, again with the TaylorF2 (7.5 PN) waveform model.

| Equation of state | $M_B^0$ ($M_\odot$) | $E_c/c^2$ ($10^{15}$ g cm$^{-3}$) | $\tilde{m}_0$ | $\Delta_{0.1}$ | Prompt collapse | $I_{max}$ ($M_\odot^3$) | $F_{max}$ (kHz) |
|---|---|---|---|---|---|---|---|
| APR4_EPP | $3.084^{+0.316}_{-0.070}$ | 1.48 | $1.019^{+0.683}_{-0.150}$ | 0.22 | Possibly | 84 | 1.89 |
| MS1_PP | $2.960^{+0.121}_{-0.026}$ | 0.53 | $-0.574^{+0.202}_{-0.044}$ | 0.17 | No | 158 | 1.03 |
| MS1B_PP | $2.973^{+0.144}_{-0.030}$ | 0.55 | $-0.542^{+0.231}_{-0.048}$ | 0.16 | No | 159 | 1.05 |
| KDE0V | $3.058^{+0.219}_{-0.053}$ | 1.80 | $1.978^{+0.582}_{-0.139}$ | 0.27 | — | 66 | 1.86 |
| SKI6 | $3.028^{+0.225}_{-0.041}$ | 1.08 | $0.922^{+0.489}_{-0.087}$ | 0.22 | — | 97 | 1.66 |
| H4 | $3.007^{+0.178}_{-0.035}$ | 0.94 | $1.450^{+0.388}_{-0.077}$ | 0.22 | Possibly | 104 | 1.36 |
| MPA1 | $3.041^{+0.255}_{-0.043}$ | 0.90 | $0.050^{+0.456}_{-0.077}$ | 0.18 | No | 121 | 1.41 |
| SK255 | $3.003^{+0.203}_{-0.041}$ | 1.10 | $1.207^{+0.497}_{-0.102}$ | 0.24 | — | 91 | 1.61 |
| SKI2 | $3.001^{+0.176}_{-0.038}$ | 0.97 | $1.100^{+0.408}_{-0.089}$ | 0.23 | — | 100 | 1.53 |
| SKI4 | $3.028^{+0.228}_{-0.037}$ | 1.11 | $0.978^{+0.494}_{-0.081}$ | 0.22 | — | 95 | 1.67 |
| SKOP | $3.019^{+0.190}_{-0.032}$ | 1.45 | $1.874^{+0.500}_{-0.085}$ | 0.26 | — | 73 | 1.70 |
| SLY2 | $3.043^{+0.239}_{-0.045}$ | 1.48 | $1.494^{+0.585}_{-0.110}$ | 0.25 | — | 77 | 1.79 |
| SLY9 | $3.018^{+0.234}_{-0.037}$ | 1.15 | $1.068^{+0.537}_{-0.083}$ | 0.23 | — | 91 | 1.68 |
| BHF_BBB2 | $3.077^{+0.201}_{-0.063}$ | 2.02 | $2.181^{+0.545}_{-0.168}$ | 0.27 | — | 63 | 1.87 |
| HQC18 | $3.064^{+0.253}_{-0.057}$ | 1.42 | $1.418^{+0.561}_{-0.129}$ | 0.22 | — | 81 | 1.72 |
| KDE0V1 | $3.049^{+0.216}_{-0.051}$ | 1.73 | $1.953^{+0.580}_{-0.135}$ | 0.27 | — | 68 | 1.80 |
| RS | $3.016^{+0.216}_{-0.042}$ | 1.14 | $1.259^{+0.507}_{-0.099}$ | 0.24 | — | 92 | 1.61 |
| SK272 | $3.003^{+0.209}_{-0.044}$ | 1.00 | $0.868^{+0.483}_{-0.102}$ | 0.23 | — | 100 | 1.61 |
| SKI3 | $3.000^{+0.190}_{-0.039}$ | 0.91 | $0.818^{+0.421}_{-0.087}$ | 0.22 | — | 107 | 1.53 |
| SKI5 | $2.975^{+0.116}_{-0.024}$ | 0.78 | $0.816^{+0.262}_{-0.056}$ | 0.23 | — | 109 | 1.47 |
| SKMP | $3.019^{+0.216}_{-0.035}$ | 1.19 | $1.254^{+0.504}_{-0.083}$ | 0.23 | — | 88 | 1.68 |
| SLY | $3.042^{+0.238}_{-0.044}$ | 1.50 | $1.517^{+0.590}_{-0.109}$ | 0.25 | Possibly | 76 | 1.82 |
| SLY230A | $3.044^{+0.251}_{-0.044}$ | 1.36 | $1.261^{+0.581}_{-0.101}$ | 0.23 | — | 83 | 1.76 |

## 5. Conclusions

In this work, we employ Bayesian model selection as a tool to discriminate between different theoretical models for the neutron star equation of state using only the gravitational-wave data. This allows us to make quantitative statements about the relative likelihood of a given equation of state model. This work complements the analysis of tidal effects of GW170817 in [6, 38] and the analysis of the neutron star radii and equation of state in [39]. As in previous studies [6, 38], we assume *a priori* that the source of GW170817 resides in NGC 4993 and adopt fixed sky location consistent with electromagnetic observations. The distance to GW170817 is determined from the gravitational-wave data alone, as in [6]. We do not allow tidal parameters to vary independently and instead enforce that each neutron star obeys a given equation of state model.

The Bayesian evidence is computed using two independent methods that show good agreement and demonstrate the robustness of our results. Under minimal assumptions on the astrophysical priors for the component masses and spins, we find that the models predicting the largest fiducial radii and tidal deformabilities are *very strongly disfavored* (MS1_PP) or at least *strongly disfavored* (MS1B_PP) with respect to the equations of state that predict a small fiducial radius and tidal deformability, such as KDE0V. In addition, we find that for both priors considered in this work the gravitational-wave data does not favor and nor can it rule out any particular equation of state against the low-mass binary black hole hypothesis.

In order to gauge waveform systematics, we compare the Bayes factors using different waveform approximants. The results show good agreement between the different waveform models, providing confidence that systematic uncertainties are small compared to statistical uncertainties.

In a broader context, we also address the question as to whether we can exclude the possibility of GW170817 being a binary black hole by computing Bayes factors against the binary neutron star and neutron star-black hole hypotheses. From the gravitational-wave data alone, no single hypothesis is strongly preferred and we are unable to exclude the binary black hole hypothesis. We find that the Bayes factors computed using different waveform approximants are in good agreement, supporting the notion that waveform systematics are small compared to statistical errors.

For some equation of state models, we can predict the outcome of the merger; for two models the remnant has to be a stable neutron star, for some it will be a long-lived supramassive neutron star, and for some a short-lived hypermassive neutron star or black hole. We can further rule out prompt collapse for a number of models. Considering the Bayes factor for each model, we find that only indefinitely stable remnants are disfavored. Based on gravitational-wave data alone, all scenarios from very long-lived remnants to prompt black hole formation at merger remain possible. Further constraints on the outcome will likely have to rely on the electromagnetic counterparts and modeling of the kilonova and the short gamma-ray burst. We find that making the assumption of a short-lived remnant—as required by some short gamma-ray burst models—would allow us to put further constraints on the equation of state model. In particular, we improve the upper limit on the maximum mass of non-rotating neutron stars, compared to an earlier study [4] (which is based on an analysis that did not assume that both neutron stars obey the same equation of state). We also find that three equation of state models are ruled out if a black hole was formed before the short gamma-ray burst, including the two disfavored by gravitational-wave data alone.

## Acknowledgments

The authors gratefully acknowledge the support of the United States National Science Foundation (NSF) for the construction and operation of the LIGO Laboratory and Advanced LIGO as well as the Science and Technology Facilities Council (STFC) of the United Kingdom, the Max-Planck-Society (MPS), and the State of Niedersachsen/Germany for support of the construction of Advanced LIGO and construction and operation of the GEO600 detector. Additional support for Advanced LIGO was provided by the Australian Research Council. The authors gratefully acknowledge the Italian Istituto Nazionale di Fisica Nucleare (INFN), the French Centre National de la Recherche Scientifique (CNRS) and the Foundation for Fundamental Research on Matter supported by the Netherlands Organisation for Scientific

Research, for the construction and operation of the Virgo detector and the creation and support of the EGO consortium. The authors also gratefully acknowledge research support from these agencies as well as by the Council of Scientific and Industrial Research of India, the Department of Science and Technology, India, the Science & Engineering Research Board (SERB), India, the Ministry of Human Resource Development, India, the Spanish Agencia Estatal de Investigación, the Vicepresidència i Conselleria d'Innovació, Recerca i Turisme and the Conselleria d'Educació i Universitat del Govern de les Illes Balears, the Conselleria d'Educació, Investigació, Cultura i Esport de la Generalitat Valenciana, the National Science Centre of Poland, the Swiss National Science Foundation (SNSF), the Russian Foundation for Basic Research, the Russian Science Foundation, the European Commission, the European Regional Development Funds (ERDF), the Royal Society, the Scottish Funding Council, the Scottish Universities Physics Alliance, the Hungarian Scientific Research Fund (OTKA), the Lyon Institute of Origins (LIO), the Paris Île-de-France Region, the National Research, Development and Innovation Office Hungary (NKFIH), the National Research Foundation of Korea, Industry Canada and the Province of Ontario through the Ministry of Economic Development and Innovation, the Natural Science and Engineering Research Council Canada, the Canadian Institute for Advanced Research, the Brazilian Ministry of Science, Technology, Innovations, and Communications, the International Center for Theoretical Physics South American Institute for Fundamental Research (ICTP-SAIFR), the Research Grants Council of Hong Kong, the National Natural Science Foundation of China (NSFC), the Leverhulme Trust, the Research Corporation, the Ministry of Science and Technology (MOST), Taiwan and the Kavli Foundation. The authors gratefully acknowledge the support of the NSF, STFC, MPS, INFN, CNRS and the State of Niedersachsen/Germany for provision of computational resources.

## Appendix A. Equations of state

In the following, we provide details for the equations of state employed in this work, all of which represent estimates from nuclear physics. They employ different estimates for the forces between constituents, for example from nucleon-nucleon scattering experiments. Those forces are often parametrized (e.g. Skyrme forces) before use in equation of state computations. Other important observable parameters to calibrate equation of state models are the nuclear symmetry energy and incompressibility. Further, the many-particle quantum field theory problem is approximately solved using different schemes, such as liquid-drop model, relativistic/non-relativistic mean field (MF) theory, Hartree–Fock (HF) method, Brueckner–Bethe–Goldstone (BG) methods, or Wigner–Seitz (WS) cells. Some methods apply the simplification of a single microstate, while others consider nuclear statistic equilibrium. One important aspect is which types of constituents are considered, for example if hyperons are included or not. Most equation of state models considered describe hadronic matter, except the HQC18 equation of state model, which assumes a hadronic crust around a quark matter core [9].

For details on the various models, we refer to the references given in table A1, which list all our equation of state models, the respective sources for the numeric tables, and the original publications. Further, we provide the maximum gravitational and baryonic masses of non-rotating neutron stars, and the maximum baryonic mass of uniformly rotating neutron stars. The latter were obtained using the Rapidly Rotating Neutron Star (RNS) code described in [126].

Most of our equations of state were obtained from three catalogs. The CompOSE project [127] provides nuclear physics equation of state data and a supporting software framework.

**Table A1.** Equations of state used in this work. We provide the original publications introducing the equations of state and the source of the numerical tables or piecewise polytropic fit parameters. For each we computed the maximum gravitational mass $\hat{M}_{\mathrm{G}}^{\mathrm{TOV}}$ of non-rotating neutron stars and the corresponding baryonic mass $\hat{M}_{\mathrm{B}}^{\mathrm{TOV}}$, as well as the maximum baryonic mass $\hat{M}_{\mathrm{B}}^{\mathrm{ROT}}$ of uniformly rotating neutron star. The WFF1 equation of state model becomes acausal below the central density of the maximum mass model, and the reported values (in brackets) are the maxima of neutron star models respecting causality.

| EoS | References | Table | Methods | $\hat{M}_{\mathrm{G}}^{\mathrm{TOV}}$ $(M_\odot)$ | $\hat{M}_{\mathrm{B}}^{\mathrm{TOV}}$ $(M_\odot)$ | $\hat{M}_{\mathrm{B}}^{\mathrm{ROT}}$ $(M_\odot)$ |
|---|---|---|---|---|---|---|
| BHF_BBB2 | [131] | [127] | Brueckner-BG | 1.92 | 2.27 | 2.64 |
| WFF1 | [98] | [128] | Variational | (1.92) | (2.31) | (2.84) |
| KDE0V | [132–134] | [127] | Skyrme,WS,HF | 1.96 | 2.31 | 2.69 |
| KDE0V1 | [132–134] | [127] | Skyrme,WS,HF | 1.97 | 2.32 | 2.69 |
| SKOP | [132, 133, 135] | [127] | Skyrme,WS,HF | 1.97 | 2.30 | 2.69 |
| H4 | [136] | [128] | Relativistic MF | 2.03 | 2.34 | 2.80 |
| HQC18 | [9] | | Hadron-Quark | 2.05 | 2.43 | 2.88 |
| SLY | [137] | [128] | Potential-method | 2.05 | 2.43 | 2.83 |
| SLY2 | [132, 133] | [127] | Skyrme,WS,HF | 2.05 | 2.43 | 2.84 |
| SLY230A | [132, 133, 138] | [127] | Skyrme,WS,HF | 2.10 | 2.50 | 2.93 |
| SKMP | [132, 133, 139] | [127] | Skyrme,WS,HF | 2.11 | 2.48 | 2.91 |
| RS | [132, 133, 140] | [127] | Skyrme,WS,HF | 2.12 | 2.48 | 2.91 |
| SK255 | [132, 133, 141] | [127] | Skyrme,WS,HF | 2.14 | 2.51 | 2.92 |
| SLY9 | [132, 133] | [127] | Skyrme,WS,HF | 2.16 | 2.55 | 2.99 |
| APR4_EPP | [129] | [130] | Variational | 2.16 | 2.61 | 3.07 |
| SKI2 | [132, 133, 142] | [127] | Skyrme,WS,HF | 2.16 | 2.53 | 2.96 |
| SKI4 | [132, 133, 142] | [127] | Skyrme,WS,HF | 2.17 | 2.58 | 3.04 |
| SKI6 | [132, 133, 143] | [127] | Skyrme,WS,HF | 2.19 | 2.60 | 3.06 |
| SK272 | [132, 133, 141] | [127] | Skyrme,WS,HF | 2.23 | 2.63 | 3.06 |
| SKI3 | [132, 133, 142] | [127] | Skyrme,WS,HF | 2.24 | 2.63 | 3.08 |
| SKI5 | [132, 133, 142] | [127] | Skyrme,WS,HF | 2.24 | 2.62 | 3.06 |
| MPA1 | [144] | [128] | Relativistic HF | 2.46 | 3.01 | 3.57 |
| MS1B_PP | [145] | [47] | Relativistic MF | 2.75 | 3.31 | 3.93 |
| MS1_PP | [145] | [47] | Relativistic MF | 2.75 | 3.30 | 3.90 |

We used this public code (selecting the first-order interpolation option and a resolution of 500 points) to create tables. Another collection of tables is provided by [128]. To obtain the final, continuous equations of state, we employ two different interpolation schemes. The results in section 4 and table A1 are computed using monotonic spline interpolation (PCHIP), while the equation of state related code used during parameter estimation employs standard cubic spline interpolation instead. For all models, we compared sequences of non-rotating neutron stars between the two. Since the low resolution and gaps present in the MS1 and MS1B tables cause significant ambiguities in the tidal deformability and radius, we use well defined piecewise polytropic approximations instead, taken from a third catalog [47] providing analytic fits, and denoted here by MS1_PP and MS1B_PP, respectively. When comparing with [6], note that their WFF1, H4, MPA1, SLY and APR4 models also refer to the piecewise polytropic variants from [47], in contrast to this work.

We applied the following adjustments: in general, we removed isolated data points from the original tables that violate thermodynamic constraints for zero temperature equations of state. For the H4 equation of state model from [128], the low-density part ($<10^{13}$ g cm$^{-3}$) was

**Table B1.** Bayes factors computed using LALINFERENCE's nested sampling algorithm for all of the equations of state discussed in section 2.4 using the narrow prior discussed in section 2.2. The Bayes factors are computed for a number of waveform models, as discussed in section 2.3. Here the Bayes factors are taken to be the ratio between the evidence for a given equation of state and the corresponding evidence for the hypothesis that both bodies are low-mass black holes. The errors correspond to statistical uncertainty in the calculation of the evidence from the nested sampling algorithm in LALINFERENCE, where $\sigma_{\mathrm{BF}} = \sigma_{\mathcal{Z}_{\mathrm{EOS}}} / \mathcal{Z}_{\mathrm{BBH}}$.

| | LALINFERENCE nest results | | | | |
|---|---|---|---|---|---|
| Equation of state | TaylorF2 (7.5pN) | Phenom+NRT | EOB+NRT | TaylorF2 (6pN) | TaylorF2 (7pN) |
| APR4_EPP | $(1.92 \pm 0.07) \times 10^{1}$ | $(4.39 \pm 0.23) \times 10^{1}$ | $(2.02 \pm 0.08) \times 10^{1}$ | $(1.76 \pm 0.06) \times 10^{1}$ | $(1.78 \pm 0.10) \times 10^{1}$ |
| BBH | $(1.00 \pm 0.05)$ | $(1.00 \pm 0.04)$ | $(1.00 \pm 0.05)$ | $(1.00 \pm 0.05)$ | $(1.00 \pm 0.05)$ |
| BHF_BBB2 | $(1.60 \pm 0.06) \times 10^{1}$ | $(4.79 \pm 0.19) \times 10^{1}$ | $(2.43 \pm 0.07) \times 10^{1}$ | $(1.70 \pm 0.07) \times 10^{1}$ | $(1.71 \pm 0.09) \times 10^{1}$ |
| H4 | $(1.73 \pm 0.08)$ | $(1.05 \pm 0.06)$ | $(3.80 \pm 0.16) \times 10^{-1}$ | $(1.73 \pm 0.08)$ | $(1.76 \pm 0.09)$ |
| HQC18 | $(1.94 \pm 0.07) \times 10^{1}$ | $(3.84 \pm 0.20) \times 10^{1}$ | $(1.95 \pm 0.12) \times 10^{1}$ | $(1.73 \pm 0.07) \times 10^{1}$ | $(1.81 \pm 0.09) \times 10^{1}$ |
| KDE0V | $(1.96 \pm 0.11) \times 10^{1}$ | $(3.95 \pm 0.17) \times 10^{1}$ | $(1.90 \pm 0.08) \times 10^{1}$ | $(1.80 \pm 0.10) \times 10^{1}$ | $(1.63 \pm 0.06) \times 10^{1}$ |
| KDE0V1 | $(1.78 \pm 0.09) \times 10^{1}$ | $(3.28 \pm 0.08) \times 10^{1}$ | $(1.54 \pm 0.07) \times 10^{1}$ | $(1.86 \pm 0.10) \times 10^{1}$ | $(1.53 \pm 0.08) \times 10^{1}$ |
| MPA1 | $(5.56 \pm 0.23)$ | $(1.35 \pm 0.05) \times 10^{1}$ | $(7.32 \pm 0.33)$ | $(4.66 \pm 0.20)$ | $(4.36 \pm 0.19)$ |
| MS1B_PP | $(3.48 \pm 0.13) \times 10^{-1}$ | $(1.61 \pm 0.07) \times 10^{-1}$ | $(7.03 \pm 0.37) \times 10^{-2}$ | $(1.86 \pm 0.07) \times 10^{-1}$ | $(1.92 \pm 0.09) \times 10^{-1}$ |
| MS1_PP | $(8.96 \pm 0.38) \times 10^{-2}$ | $(8.45 \pm 0.46) \times 10^{-2}$ | $(3.36 \pm 0.14) \times 10^{-2}$ | $(3.44 \pm 0.16) \times 10^{-2}$ | $(3.44 \pm 0.14) \times 10^{-2}$ |
| RS | $(3.66 \pm 0.15)$ | $(9.85 \pm 0.46)$ | $(3.73 \pm 0.17)$ | $(3.78 \pm 0.15)$ | $(4.05 \pm 0.15)$ |
| SK255 | $(3.75 \pm 0.19)$ | $(9.56 \pm 0.35)$ | $(3.79 \pm 0.17)$ | $(3.77 \pm 0.16)$ | $(3.75 \pm 0.15)$ |
| SK272 | $(3.22 \pm 0.11)$ | $(5.54 \pm 0.22)$ | $(2.74 \pm 0.11)$ | $(3.65 \pm 0.17)$ | $(3.55 \pm 0.20)$ |
| SKI2 | $(2.07 \pm 0.10)$ | $(4.06 \pm 0.18)$ | $(1.63 \pm 0.05)$ | $(2.13 \pm 0.10)$ | $(2.69 \pm 0.14)$ |
| SKI3 | $(2.03 \pm 0.08)$ | $(3.50 \pm 0.13)$ | $(1.46 \pm 0.08)$ | $(2.23 \pm 0.07)$ | $(2.54 \pm 0.06)$ |
| SKI4 | $(7.35 \pm 0.32)$ | $(1.35 \pm 0.06) \times 10^{1}$ | $(7.66 \pm 0.22)$ | $(4.99 \pm 0.22)$ | $(4.32 \pm 0.19)$ |
| SKI5 | $(1.15 \pm 0.05)$ | $(3.40 \pm 0.16) \times 10^{-1}$ | $(1.38 \pm 0.05) \times 10^{-1}$ | $(6.39 \pm 0.26) \times 10^{-1}$ | $(7.08 \pm 0.29) \times 10^{-1}$ |
| SKI6 | $(6.38 \pm 0.28)$ | $(1.42 \pm 0.07) \times 10^{1}$ | $(6.67 \pm 0.22)$ | $(4.34 \pm 0.20)$ | $(4.24 \pm 0.20)$ |
| SKMP | $(6.80 \pm 0.26)$ | $(1.46 \pm 0.07) \times 10^{1}$ | $(7.32 \pm 0.37)$ | $(4.43 \pm 0.16)$ | $(4.15 \pm 0.17)$ |
| SKOP | $(1.50 \pm 0.07) \times 10^{1}$ | $(1.02 \pm 0.06) \times 10^{1}$ | $(4.72 \pm 0.14)$ | $(9.46 \pm 0.42)$ | $(8.79 \pm 0.39)$ |

*(Continued)*

**Table B1.** (*Continued*)

| Equation of state | TaylorF2 (7.5pN) | Phenom+NRT | EOB+NRT | TaylorF2 (6pN) | TaylorF2 (7pN) |
|---|---|---|---|---|---|
| SLY | $(1.85 \pm 0.09) \times 10^1$ | $(1.82 \pm 0.09) \times 10^1$ | $(6.69 \pm 0.29)$ | $(1.55 \pm 0.06) \times 10^1$ | $(1.39 \pm 0.06) \times 10^1$ |
| SLY2 | $(1.68 \pm 0.05) \times 10^1$ | $(2.00 \pm 0.07) \times 10^1$ | $(7.90 \pm 0.31)$ | $(1.53 \pm 0.07) \times 10^1$ | $(1.40 \pm 0.09) \times 10^1$ |
| SLY230A | $(1.80 \pm 0.07) \times 10^1$ | $(1.47 \pm 0.05) \times 10^1$ | $(6.26 \pm 0.28)$ | $(1.42 \pm 0.06) \times 10^1$ | $(1.18 \pm 0.04) \times 10^1$ |
| SLY9 | $(8.93 \pm 0.43)$ | $(1.42 \pm 0.06) \times 10^1$ | $(8.11 \pm 0.28)$ | $(5.46 \pm 0.23)$ | $(4.76 \pm 0.20)$ |
| WFF1 | $(1.06 \pm 0.05) \times 10^1$ | $(3.49 \pm 0.16) \times 10^1$ | $(2.45 \pm 0.11) \times 10^1$ | $(1.27 \pm 0.05) \times 10^1$ | $(1.34 \pm 0.06) \times 10^1$ |

**Table B2.** Similarly to table B1, we show the Bayes factors computed using LALInference's nested sampling algorithm for all of the equations of state discussed in section 2.4 using the broad prior discussed in section 2.2.

| | LALInference nest results | | |
|---|---|---|---|
| Equation of state | TaylorF2 (7.5pN) | Phenom+NRT | EOB+NRT |
| APR4_EPP | $(3.21 \pm 0.13)$ | $(2.56 \pm 0.08)$ | $(1.40 \pm 0.06)$ |
| BBH | $(1.00 \pm 0.05)$ | $(1.00 \pm 0.07)$ | $(1.00 \pm 0.06)$ |
| BHF_BBB2 | $(3.42 \pm 0.18)$ | $(3.23 \pm 0.17)$ | $(1.86 \pm 0.14)$ |
| H4 | $(1.88 \pm 0.10) \times 10^{-1}$ | $(5.21 \pm 0.42) \times 10^{-2}$ | $(2.86 \pm 0.10) \times 10^{-2}$ |
| HQC18 | $(2.97 \pm 0.16)$ | $(2.06 \pm 0.10)$ | $(1.27 \pm 0.06)$ |
| KDE0V | $(3.12 \pm 0.13)$ | $(1.98 \pm 0.12)$ | $(1.51 \pm 0.10)$ |
| KDE0V1 | $(2.88 \pm 0.13)$ | $(1.71 \pm 0.09)$ | $(9.97 \pm 0.41) \times 10^{-1}$ |
| MPA1 | $(6.15 \pm 0.34) \times 10^{-1}$ | $(8.30 \pm 0.36) \times 10^{-1}$ | $(5.54 \pm 0.41) \times 10^{-1}$ |
| MS1B_PP | $(3.65 \pm 0.19) \times 10^{-2}$ | $(9.71 \pm 0.38) \times 10^{-3}$ | $(6.94 \pm 0.36) \times 10^{-3}$ |
| MS1_PP | $(9.68 \pm 0.45) \times 10^{-3}$ | $(4.56 \pm 0.23) \times 10^{-3}$ | $(2.80 \pm 0.15) \times 10^{-3}$ |
| RS | $(3.89 \pm 0.18) \times 10^{-1}$ | $(3.43 \pm 0.20) \times 10^{-1}$ | $(2.51 \pm 0.14) \times 10^{-1}$ |
| SK255 | $(4.00 \pm 0.27) \times 10^{-1}$ | $(3.57 \pm 0.22) \times 10^{-1}$ | $(2.49 \pm 0.19) \times 10^{-1}$ |
| SK272 | $(3.44 \pm 0.15) \times 10^{-1}$ | $(2.63 \pm 0.22) \times 10^{-1}$ | $(1.99 \pm 0.12) \times 10^{-1}$ |
| SKI2 | $(2.05 \pm 0.09) \times 10^{-1}$ | $(1.51 \pm 0.13) \times 10^{-1}$ | $(8.66 \pm 0.50) \times 10^{-2}$ |
| SKI3 | $(2.09 \pm 0.09) \times 10^{-1}$ | $(1.45 \pm 0.12) \times 10^{-1}$ | $(8.26 \pm 0.53) \times 10^{-2}$ |
| SKI4 | $(8.19 \pm 0.42) \times 10^{-1}$ | $(8.30 \pm 0.57) \times 10^{-1}$ | $(6.17 \pm 0.30) \times 10^{-1}$ |
| SKI5 | $(9.59 \pm 0.41) \times 10^{-2}$ | $(1.48 \pm 0.07) \times 10^{-2}$ | $(1.35 \pm 0.08) \times 10^{-2}$ |
| SKI6 | $(6.02 \pm 0.33) \times 10^{-1}$ | $(7.97 \pm 0.58) \times 10^{-1}$ | $(5.66 \pm 0.44) \times 10^{-1}$ |
| SKMP | $(6.59 \pm 0.33) \times 10^{-1}$ | $(8.27 \pm 0.69) \times 10^{-1}$ | $(6.04 \pm 0.45) \times 10^{-1}$ |
| SKOP | $(1.47 \pm 0.07)$ | $(8.09 \pm 0.36) \times 10^{-1}$ | $(6.70 \pm 0.51) \times 10^{-1}$ |
| SLY | $(2.32 \pm 0.12)$ | $(9.53 \pm 0.59) \times 10^{-1}$ | $(6.93 \pm 0.42) \times 10^{-1}$ |
| SLY2 | $(2.20 \pm 0.11)$ | $(1.09 \pm 0.08)$ | $(6.55 \pm 0.49) \times 10^{-1}$ |
| SLY230A | $(2.19 \pm 0.13)$ | $(8.02 \pm 0.50) \times 10^{-1}$ | $(6.40 \pm 0.38) \times 10^{-1}$ |
| SLY9 | $(7.64 \pm 0.34) \times 10^{-1}$ | $(9.46 \pm 0.49) \times 10^{-1}$ | $(6.85 \pm 0.33) \times 10^{-1}$ |
| WFF1 | $(3.75 \pm 0.21)$ | $(4.32 \pm 0.25)$ | $(3.01 \pm 0.19)$ |

missing and filled in using the corresponding values of the H3 equation of state model from the same source. For the equations of state from [129], the sound speed becomes superluminal shortly before the maximum mass model is reached, both for the tabulated versions [128] (AP4) or [127] (APR), and the piecewise polytropic fit from [47] (APR4). We therefore use the modified APR4 (denoted APR4_EPP here) from [130], which adds additional segments at high density to ensure causality.

**Table B3.** Bayes factors computed using the RIFT algorithm for all of the equations of state discussed in section 2.4 using the narrow prior discussed in section 2.2. Bayes factors are computed for the costly waveform models discussed in section 2.3. Here the Bayes factors are shown as a ratio between the evidence for the given equation of state model and the corresponding evidence for the hypothesis that both bodies are neutron stars, described by the SLY9 equation of state.

| | RIFT results | |
|---|---|---|
| Equation of state | SEOBNRv4T (narrow) | TEOBResumS (narrow) |
| APR4_EPP | $(2.37 \pm 0.06)$ | $(1.46 \pm 0.04)$ |
| BBH | $(1.35 \pm 0.03)$ | $(2.76 \pm 0.06) \times 10^{-1}$ |
| BHF_BBB2 | $(2.53 \pm 0.06)$ | $(1.38 \pm 0.05)$ |
| H4 | $(1.06 \pm 0.02) \times 10^{-1}$ | $(8.26 \pm 0.18) \times 10^{-2}$ |
| HQC18 | $(2.38 \pm 0.06)$ | $(1.46 \pm 0.04)$ |
| KDE0V | $(2.36 \pm 0.05)$ | $(1.49 \pm 0.05)$ |
| KDE0V1 | $(2.30 \pm 0.06)$ | $(1.47 \pm 0.04)$ |
| MPA1 | $(7.43 \pm 0.14) \times 10^{-1}$ | $(7.64 \pm 0.19) \times 10^{-1}$ |
| MS1B_PP | $(2.59 \pm 0.06) \times 10^{-2}$ | $(1.22 \pm 0.02) \times 10^{-2}$ |
| MS1_PP | $(6.98 \pm 0.19) \times 10^{-3}$ | $(2.99 \pm 0.07) \times 10^{-3}$ |
| RS | $(4.43 \pm 0.08) \times 10^{-1}$ | $(4.61 \pm 0.11) \times 10^{-1}$ |
| SK255 | $(4.61 \pm 0.09) \times 10^{-1}$ | $(4.90 \pm 0.12) \times 10^{-1}$ |
| SK272 | $(3.42 \pm 0.07) \times 10^{-1}$ | $(3.41 \pm 0.08) \times 10^{-1}$ |
| SKI2 | $(1.67 \pm 0.03) \times 10^{-1}$ | $(1.55 \pm 0.03) \times 10^{-1}$ |
| SKI3 | $(1.64 \pm 0.03) \times 10^{-1}$ | $(1.43 \pm 0.03) \times 10^{-1}$ |
| SKI4 | $(9.07 \pm 0.22) \times 10^{-1}$ | $(9.27 \pm 0.21) \times 10^{-1}$ |
| SKI5 | $(5.95 \pm 0.13) \times 10^{-2}$ | $(3.69 \pm 0.07) \times 10^{-2}$ |
| SKI6 | $(8.05 \pm 0.16) \times 10^{-1}$ | $(8.32 \pm 0.21) \times 10^{-1}$ |
| SKMP | $(8.50 \pm 0.17) \times 10^{-1}$ | $(8.73 \pm 0.23) \times 10^{-1}$ |
| SKOP | $(1.51 \pm 0.03)$ | $(1.29 \pm 0.03)$ |
| SLY | $(1.92 \pm 0.04)$ | $(1.42 \pm 0.04)$ |
| SLY2 | $(1.96 \pm 0.04)$ | $(1.46 \pm 0.04)$ |
| SLY230A | $(1.80 \pm 0.04)$ | $(1.42 \pm 0.04)$ |
| SLY9 | $(1.00 \pm 0.02)$ | $(1.00 \pm 0.02)$ |
| WFF1 | $(2.67 \pm 0.06)$ | $(1.16 \pm 0.03)$ |

## Appendix B. Bayes factors

Here we present the explicit values for the Bayes factors computed using the nested sampling algorithm implemented in LALInference and RIFT. We show results for all of the equations of state discussed in section 2.4 for both the wide and narrow priors outlined in section 2.2.

**Table B4.** Bayes factors computed using the RIFT algorithm for all of the equations of state discussed in section 2.4 using the broad prior discussed in section 2.2. Bayes factors are computed for the costly waveform models discussed in section 2.3. Here the Bayes factors are shown as a ratio between the evidence for the given equation of state model and the corresponding evidence for the hypothesis that both bodies are neutron stars, described by the SLY9 equation of state.

| | RIFT results | |
|---|---|---|
| Equation of state | SEOBNRv4T (broad) | TEOBResumS (broad) |
| APR4_EPP | $(2.67 \pm 0.08)$ | $(1.64 \pm 0.05)$ |
| BBH | $(1.22 \pm 0.04)$ | $(2.06 \pm 0.09) \times 10^{-1}$ |
| BHF_BBB2 | $(2.89 \pm 0.08)$ | $(1.65 \pm 0.06)$ |
| H4 | $(2.33 \pm 0.08) \times 10^{-1}$ | $(1.79 \pm 0.06) \times 10^{-1}$ |
| HQC18 | $(2.51 \pm 0.08)$ | $(1.69 \pm 0.05)$ |
| KDE0V | $(2.50 \pm 0.09)$ | $(1.70 \pm 0.05)$ |
| KDE0V1 | $(2.39 \pm 0.07)$ | $(1.69 \pm 0.06)$ |
| MPA1 | $(7.85 \pm 0.25) \times 10^{-1}$ | $(8.06 \pm 0.19) \times 10^{-1}$ |
| MS1B_PP | $(4.82 \pm 0.18) \times 10^{-2}$ | $(3.88 \pm 0.20) \times 10^{-2}$ |
| MS1_PP | $(5.42 \pm 0.23) \times 10^{-3}$ | $(3.47 \pm 0.18) \times 10^{-3}$ |
| RS | $(4.91 \pm 0.15) \times 10^{-1}$ | $(4.54 \pm 0.11) \times 10^{-1}$ |
| SK255 | $(4.99 \pm 0.14) \times 10^{-1}$ | $(4.66 \pm 0.12) \times 10^{-1}$ |
| SK272 | $(4.18 \pm 0.14) \times 10^{-1}$ | $(3.63 \pm 0.10) \times 10^{-1}$ |
| SKI2 | $(2.91 \pm 0.09) \times 10^{-1}$ | $(2.18 \pm 0.06) \times 10^{-1}$ |
| SKI3 | $(2.78 \pm 0.09) \times 10^{-1}$ | $(2.18 \pm 0.07) \times 10^{-1}$ |
| SKI4 | $(9.19 \pm 0.26) \times 10^{-1}$ | $(9.53 \pm 0.28) \times 10^{-1}$ |
| SKI5 | $(1.48 \pm 0.05) \times 10^{-1}$ | $(1.17 \pm 0.05) \times 10^{-1}$ |
| SKI6 | $(8.22 \pm 0.24) \times 10^{-1}$ | $(8.38 \pm 0.21) \times 10^{-1}$ |
| SKMP | $(8.21 \pm 0.28) \times 10^{-1}$ | $(8.36 \pm 0.23) \times 10^{-1}$ |
| SKOP | $(1.50 \pm 0.05)$ | $(1.38 \pm 0.04)$ |
| SLY | $(2.01 \pm 0.06)$ | $(1.62 \pm 0.05)$ |
| SLY2 | $(1.97 \pm 0.07)$ | $(1.61 \pm 0.05)$ |
| SLY230A | $(1.87 \pm 0.06)$ | $(1.56 \pm 0.05)$ |
| SLY9 | $(1.00 \pm 0.03)$ | $(1.00 \pm 0.03)$ |
| WFF1 | $(2.89 \pm 0.08)$ | $(1.20 \pm 0.04)$ |